\documentclass[11pt]{article}
\usepackage{jheppub,relsize}

\usepackage{slashed,ulem}
\usepackage{graphicx}
\usepackage{subfigure}
\usepackage{dcolumn}
\usepackage{bm}
\usepackage{float}
\usepackage{amsfonts}
\usepackage{xcolor}

\newcommand{\be}{\begin{equation}} 
\newcommand{\ee}{\end{equation}}  
\newcommand{\bea}{\begin{eqnarray}}  
\newcommand{\eea}{\end{eqnarray}}

\begin{document}

\title{Exotic Gravitational Wave Signatures from Simultaneous Phase Transitions}

\author{Djuna Croon,}\affiliation{Department of Physics and Astronomy, Dartmouth College, Hanover, NH 03755, USA}
\author{Graham White}%
\affiliation{TRIUMF Theory Group, 4004 Wesbrook Mall, Vancouver, B.C. V6T2A3, Canada}
\date{\today}

\abstract{We demonstrate that the relic gravitational wave background from a multi-step phase transition may deviate from the simple sum of the single spectra, for phase transitions with similar nucleation temperatures $T_N$. We demonstrate that the temperature range $\Delta T$ between the volume fractions $f(T)=0.1$ and $f(T)=0.9$ occupied by the vacuum bubbles can span $\sim 20$ GeV. This allows for a situation in which phase transitions overlap, such that the later bubbles may nucleate both in high temperature and intermediate temperature phases. Such scenarios may lead to more exotic gravitational wave spectra, which cannot be fitted that of a consecutive PTs. We demonstrate this explicitly in the singlet extension of the Standard Model. Finally, we comment on potential additional effects due to the more exotic dynamics of overlapping phase transitions.}
\maketitle

\section{Introduction}

The detection of gravitational waves (GW) \cite{Abbott:2016blz} established an important probe of New Physics. Deviations in the waveform of resolvable events could point to interacting dark matter \cite{Ellis:2017jgp,Croon:2017zcu}, exotic compact objects \cite{Giudice:2016zpa,Palenzuela:2017kcg,Croon:2018ftb} or the fracturing of condensates \cite{Kusenko:2008zm,Kusenko:2009cv}.
The observation also implies that the detection of a stochastic GW background will be within reach of current and future detectors. This may be a rare probe of the Cosmic Dark Ages and the first observational window onto cosmic phase transitions (PTs). Such a cosmic phase transition leaves behind a striking relic gravitational wave spectrum that depends on the strength of the transition, the speed of the transition, the bubble wall velocity and the temperature of the transition. The electroweak phase transition in particular is of interest since a strongly first order electroweak phase transition can catalyze electroweak baryogenesis \cite{White:2016nbo,Morrissey:2012db,Trodden:1998ym,Dev:2016feu} providing an explanation for matter anti-matter asymmetry observed today \cite{deVries:2017ncy,Balazs:2016yvi,Akula:2017yfr,Balazs:2013cia,Balazs:2004ae,Lee:2004we,Morrissey:2012db,Trodden:1998ym}. The phenomenology of such PTs have been studied abundantly \cite{Balazs:2016tbi,Jinno:2017ixd,Matsui:2017ggm,Huang:2017kzu,Baldes:2017ygu,Demidov:2017lzf,Chala:2018ari,Hashino:2018zsi,Kobakhidze:2016mch,Arunasalam:2017ajm} and the GW production has recently been advanced theoretically \cite{Hindmarsh:2016lnk} and through lattice simulations \cite{Hindmarsh:2017gnf,Cutting:2018tjt}.

Recently there has been much interest in multi-step phase transitions \cite{Patel:2013zla,Inoue:2015pza,Akula:2017yfr,Vieu:2018zze,Ramsey-Musolf:2017tgh,Kang:2017mkl,Patel:2012pi,Blinov:2015sna}. The electroweak phase transition can occur at a higher scale, if the symmetry breaks through a two step transition \cite{Inoue:2015pza}. Baryogenesis can occur through exotic high temperature color breaking transitions \cite{Ramsey-Musolf:2017tgh} and if a singlet acquires a vev before the electroweak phase transition, it can catalyze a strongly first order electroweak phase transition \cite{Profumo:2014opa} as well as affecting the amount of baryon asymmetry produced in the NMSSM \cite{Akula:2017yfr}. 
Here we will consider the GW signal from multi-step phase transitions, and argue that the phenomenology may be altered if the transitions overlap.
To show this, here we will compare the different situations,
\begin{itemize} \label{situations}
\item Isolated phase transitions; phase transitions which are separated in temperature from other PT. For sequential isolated PTs, the GW signal is expected to be the sum of both spectra.
\item Simultaneous phase transitions; the phase transitions overlap\footnote{We will refer to this class as simultaneous PTs even though the transitions do not have to start and end at the same temperature. More concretely, $T_f^{A} < T_N^{(B)} < T_N^{(B)}$ for PT A and B.} and bubbles of both vaccua coexist for some range $\Delta T$. As we will argue here, the GW signal is expected to deviate from the sum of single spectra, as the evolution of the bubbles may be altered.
\end{itemize}

As an example, we will study a common multi-step PT in a singlet scalar boson extension of the Standard Model.
Such new bosons are ubiquitous in extensions to the Standard Model  \cite{Chang:2017ynj,Chang:2012ta,Ng:2015eia,Ellwanger:2010oug,Athron:2009bs,Balazs:2013cia,Akula:2017yfr,Ivanov:2017dad,Fink:2018mcz,Vieu:2018nfq,Croon:2015naa,Aydemir:2015nfa,Aydemir:2016xtj,Croon:2015wba}. New scalar particles can be a dark matter candidate, the inflaton \cite{Croon:2013ana}, a portal to the dark sector \cite{Chang:2016pya,Chang:2014lxa,Cabral-Rosetti:2017mai,Garg:2017iva,McKay:2017iur,Athron:2017kgt,Burgess:2000yq,McDonald:1993ex,Cline:2013gha}, as well as a catalyzing a strongly first order electroweak phase transition \cite{Profumo:2014opa,Profumo:2007wc,White:2016nbo,Kozaczuk:2014kva} and improving the stability of the vacuum \cite{EliasMiro:2012ay,Gonderinger:2009jp,Gonderinger:2012rd,Khan:2014kba,Balazs:2016tbi}. GW detection provides a complimentary probe to this model to current and future colliders \cite{Chang:2018pjp,Contino:2016spe}. 

This paper is organized as follows. In the next section, we will give a brief review of the parameters that influence the evolution of the nucleation rate and growth of vacuum bubbles. In section \ref{multi-step}, we will describe the different nucleation cases, and show the effects in a toy model example. In section \ref{singletexample}, we will provide a more realistic application of the coexisting PT, in a singlet extension of the Standard Model. We will finish with a summary and an outlook to the experimental prospects of these effects.

\section{Thermal parameters for a cosmic phase transition}
\subsection{Review of important parameters}
For a system with multiple scalar fields evolving with temperature during a cosmic phase transition, many thermal properties are controlled by the euclidean action
\begin{equation}
    S_{\rm E} = 4\pi \int r^2 dr \left[\sum_i \frac{1}{2} \left( \frac{\partial \phi _i}{\partial r} \right)^2 +V(\phi _i ,T) \right]
\end{equation}
where we have assumed spherical symmetry for the system.
The extremum of the euclidean action that describes the formation of a bubble of a new phase forming in a vacuum of high temperature phase is called the bounce \cite{Coleman:1977py,Callan:1977pt,Linde:1980tt}. 
The bounce is a solution to the classical equations of motion
\begin{equation}
    \frac{\partial ^2 \phi _i}{\partial r^2} +\frac{2}{r} \frac{\partial \phi _i}{\partial r} = \frac{\partial V}{\partial \phi _i} 
\end{equation}
with boundary conditions $\phi _i (0) \sim \phi _i(v_i)$, $\phi _i^\prime = 0$ and $\phi _i (\infty) = (v_i^0)$ where $v_i$ and $v_i^0$ are the true and false vacua respectively. The transition temperature $T_N$ is defined as the temperature when at least one critical bubble forms within a Hubble volume and is approximately given by
\begin{equation}
    \frac{S_E}{T_N} \sim 177 -4 \ln  \frac{T_N}{\rm GeV} -2 \ln g_*  \label{eqn:TN}
\end{equation}
with some uncertainty in the initial numeric factor on the right hand side.
The characteristic rate of bubble nucleation $\beta$ is given by 
\begin{equation}
    \frac{\beta}{H} = T \frac{\partial }{\partial T } \left( \frac{S_E}{T} \right) \ .
\end{equation}
A quantity related to the latent heat of the system, $\Upsilon$, is also relevant to the relic gravitational wave background and is defined as
\begin{equation}
    \Upsilon = \frac{\rho }{\rho ^*}
\end{equation}
where $\rho^*$ is the energy density for a radiation dominated universe, and $\rho$ the difference in energy density in the vacuua on either sides of the bubble wall, 
\begin{eqnarray}
\rho ^* &=& \frac{g_* \pi ^2  T^4}{30} \\
\rho &=& \left. \left[V- \frac{\partial V}{\partial T}T \right]\right|_{v_i^0}-\left. \left[V- \frac{\partial V}{\partial T}T \right]\right|_{v_i} \ .
\end{eqnarray}
The wall velocity can also be calculated from the finite temperature effective potential $V(T,\phi^i)$ \cite{White:2016nbo}.
In the slow and fast wall regimes, the wall velocity can be calculated semi-analytically. However, for the case of Standard Model plus singlet the usual approximations for calculating the wall velocity \cite{White:2016nbo} can break down, making the calculation of the wall velocity a formidable task \cite{Kozaczuk:2015owa}. We refer the reader to \cite{Kozaczuk:2015owa} for an explicit numerical computation. Here we note that a typical wall velocity is $0.2 \leq v_w \leq 0.6$ for the electroweak transition, but may vary from $v_w \sim 0.01$ and $v_w \rightarrow 1$ depending on the medium in which the bubble expands. Specifically, the wall velocity is controlled by ``friction terms'' provided by the particles which change rest mass in the PT. The amount of friction depends on the magnitude of the mass change, and the distribution of such particles in the plasma. In particular, the wall velocity may reach ultra-relavistic values for a PT in a vacuum \cite{Kosowsky:1991ua}.

\begin{figure}
    \centering
    \includegraphics[width=.55\textwidth]{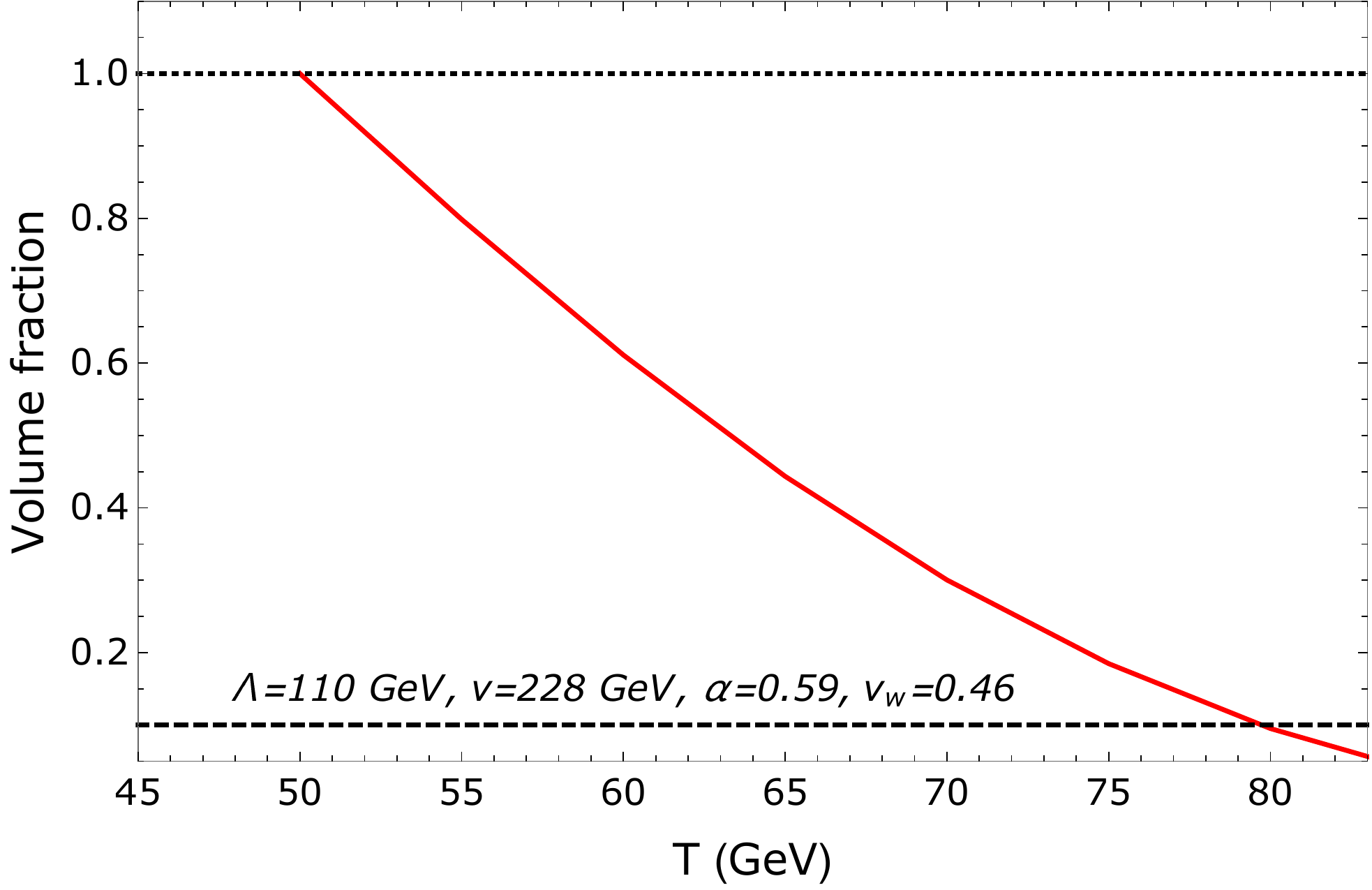}
    \caption{Benchmark example for the volume fraction \eqref{Volumefrac} as a function of temperature. For this benchmark, $T_c = 196 \text{ GeV}$ and $v(T_c) = 225 \text{ GeV}$.}
    \label{benchmark1}
\end{figure}

The final thing we will need is the growth of the volume fraction as a function of time. First, the transition rate is \cite{Linde:1980tt}
\begin{equation}
 \Gamma = T^4 \left( \frac{S_E}{2 \pi T} \right) ^{3/2}  e^{-S_E/T}  \ .  
\end{equation}
From this we can calculate the number of bubbles in a Hubble volume \cite{Anderson:1991zb,Caprini:2011uz} 
\begin{equation}
    N(T)=\int _T^{T_C} \frac{dT}{T} \frac{\Gamma }{H(T)^4} 
\end{equation}
with the Hubble parameter in equilibrium,
\begin{equation}
    H(T) = \left[\frac{8 \pi g_* T^4}{90 m_{\rm pl} ^2} \right]^{1/2}
\end{equation}
Then, we can write the new phase volume fraction as
\begin{eqnarray}\label{Volumefrac}
    f(T)&=& N(T) \times \frac{4\pi}{3}  \frac{ \left(R_c +  v_w t \right)^3}{ H(T)^{-3}}\\
        &=& N(T) \times \frac{4\pi}{3} \left(R_c +  v_w\left[ \frac{1}{T^2}-\frac{1}{T_N^2} \right]\sqrt{\frac{45}{16 \pi ^3}} \frac{m_{\rm pl}}{g_*} \right)^3 H(T)^3
\end{eqnarray}
Note that in the above we have $R_c=\delta (T) v_s(T)/\Lambda^2(T)$ in the notation for the singlet PT (in section \ref{singletexample}). 
We show an example of the evolution of $f(T)$ in Fig. \ref{benchmark1}. 
The PT finishes at $T_f$, which is defined by
\be\label{Tf} f(T_f) = 1 \ee
The duration of the PT is then given by $\Delta T = T_f-T_N $, and can be up to $\mathcal{O}(10 \,\text{GeV})$, as is seen from Fig.\ref{benchmark1}. As is obvious, simultaneous PT are more likely for relatively large $\Delta T$. 

Finally, we will comment on the duration of the PT with respect to reheating and the Hubble parameter. 
For a radiation dominated Universe, the temperature and time are related by 
\begin{equation}
    T^2 t = \sqrt{\frac{45}{16 \pi ^3}} \frac{M_p}{g_*} \ .
\end{equation}
In a more general scenario the relationship between temperature and time is more involved, as the production of bubbles releases latent heat which reheats the vanilla vacuum. The amount of reheating decreases with $x=g_l/g_*$ where $g_l$ is the number of degrees of freedom that don't acquire a mass through the phase transition \cite{Megevand:2007sv}. However, in the case of the singlet phase transition $g_l \sim g_*$, and the reheating is therefore small. Furthermore, since the bubble wall velocities are assumed to be relatively fast for our analysis, we do not expect the latent heat to diffuse throughout the vanilla vacuum. Therefore we will assume that the relevant temperature range in the vanilla vacuum will not be affected by reheating, for the majority of the volume.
We note that for $\Delta T= 10$ GeV, the duration of the PT is $\sim 10^{-10} $ s. Since this is much shorter than the Hubble scale at the time of the PT, $H_*^{-1}$, we ignore Hubble expansion in the following. For an analysis where this assumption is relaxed see \cite{Kobakhidze:2017mru}

\subsection{Coexisting bubbles: $T_N^{(1)}>T_N ^{(2)}> T_f^{(1)}$}
\begin{figure}
    \centering
    \includegraphics[width=0.35\textwidth]{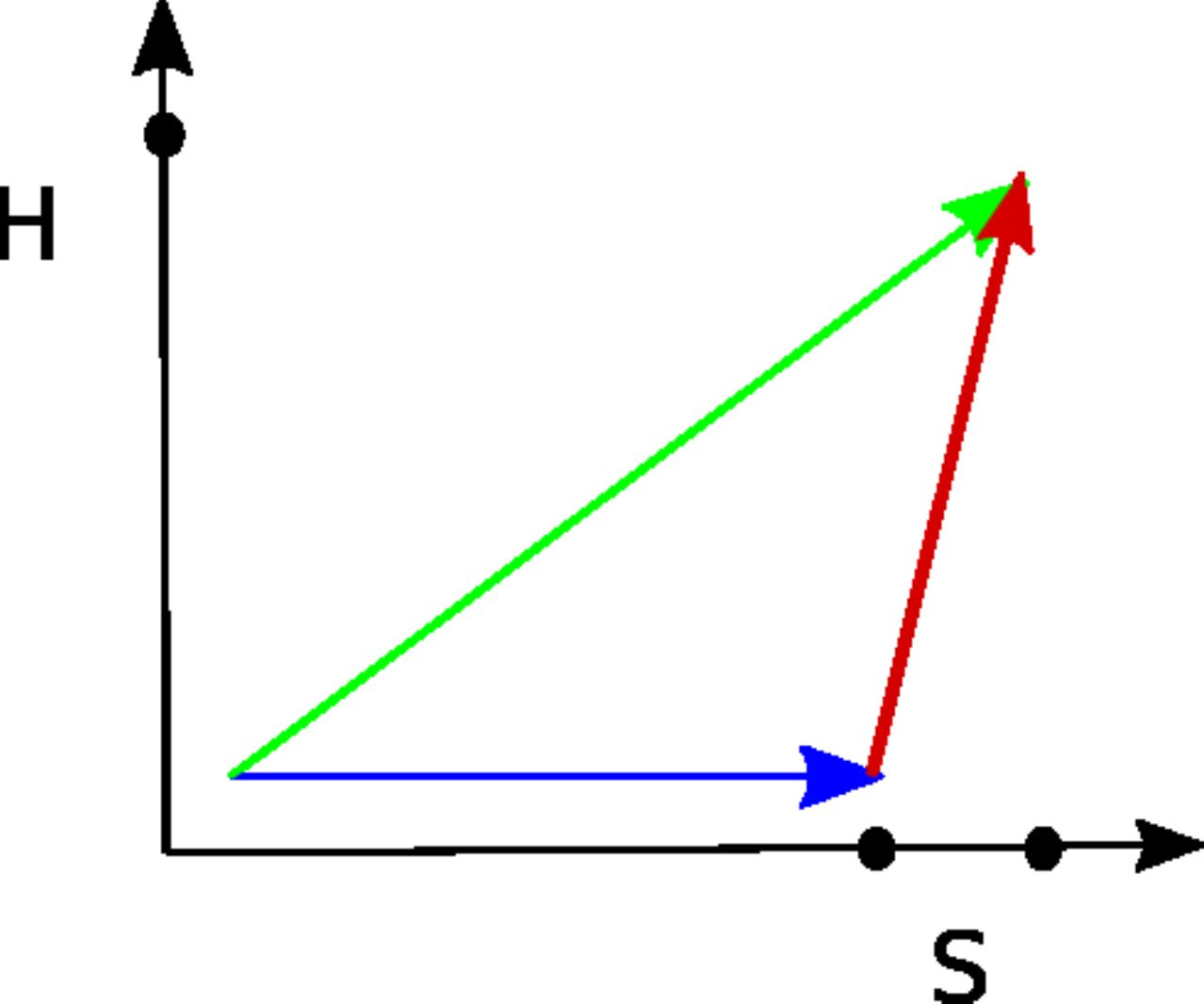}
    \includegraphics[width=0.45\textwidth]{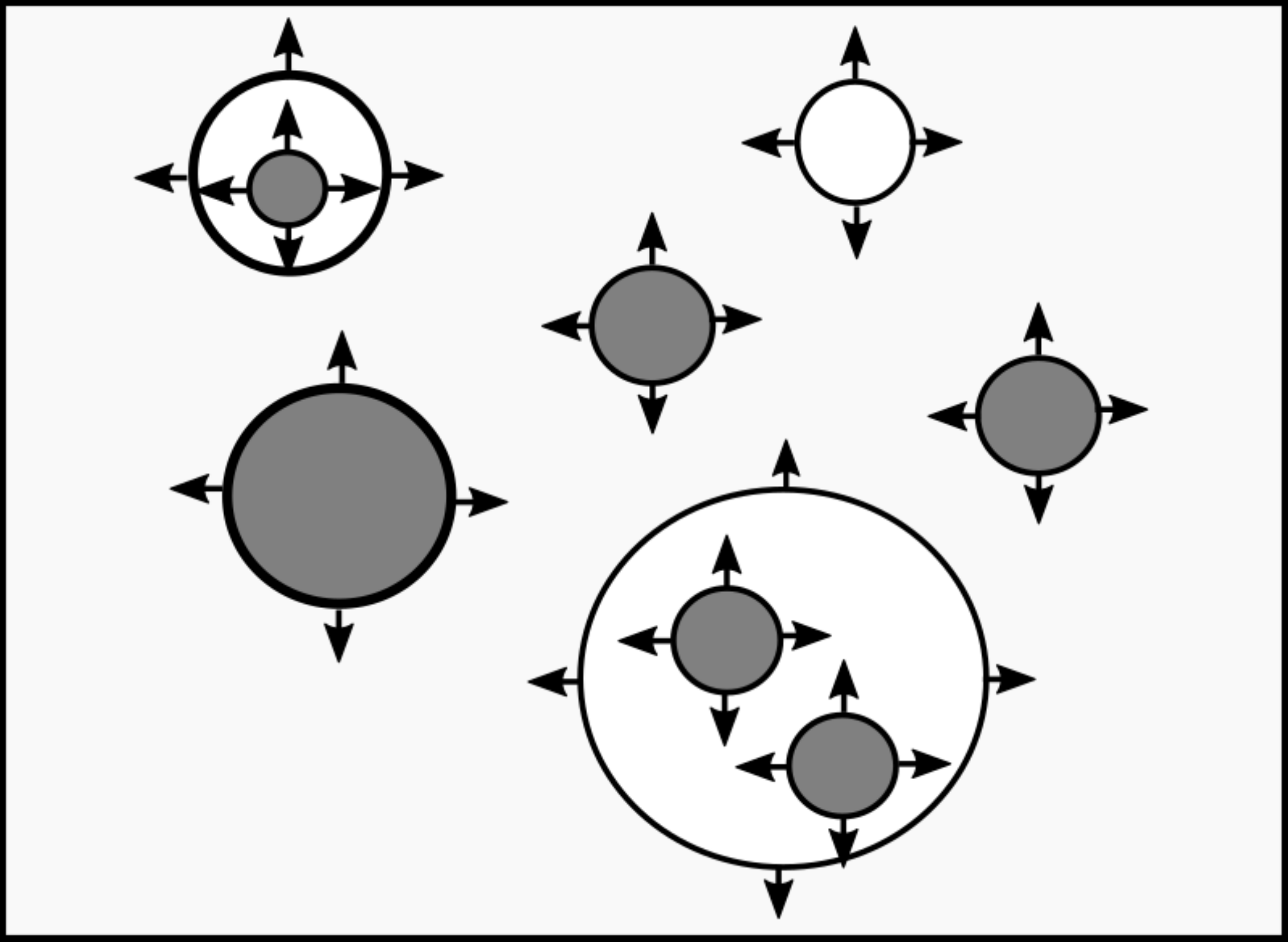}
    \caption{Left: the possible phase transitions for the case where the Standard Model is extended by a singlet. If the phase transitions are separated then only either bubbles for the blue and red transitions form. If the phase transitions are overlapping the green can form too. Right: Example of the nucleation of different bubble species when an overlapping two step phase transition occurs. The background is the vanilla vacuum and different coloured bubbles represent the vacua A and B, corresponding to minima which are deeper than the vanilla vacuum.}\label{fig:doublebubble}
\end{figure}
Now we will consider the case where the Standard Model is extended by a singlet scalar field. The vacuum transitions that can occur in a two step transition are shown in the left panel of Fig. \ref{fig:doublebubble}. 
In the case where $T_N^{(1)}>T_N ^{(2)}> T_f^{(1)}$, bubbles for the first transition may form in the vanilla vacuum, and bubbles of the true vacuum may form in both the singlet and vanilla phases as shown in the right panel of Fig. \ref{fig:doublebubble}. This scenario can occur when there is a zero temperature barrier between the true and false vacuum in the singlet sector. In this case, $S_E/T$ decreases until there is a minimum, followed by a period in which then $S_E/T$ increases as the universe cools.  Examples of this are shown in Fig. \ref{fig:thermalparameters}\footnote{In the preparation of this manuscript, the existence of such a minimum was also pointed out by \cite{Vieu:2018zze}.}. In these cases, where $S_E/T$ only drops slightly below the critical value (defining the nucleation temperature), the difference between $T_N$ and $T_f$ can be tens of GeV, as shown for an explicit example in Fig. \ref{benchmark1}. Here, the $\Delta T$ taken for the volume fraction of the new phase to grow from $0.1$ to $1$ is $\sim 20$ GeV.

Eventually the temperature is low enough so that the second species of bubble forms. If the second bubble forms within the first bubble, it will have a different transition temperature, latent heat, transition speed and wall velocity compared to the case where it forms in the vanilla vacuum. In the following, we will denote the latter as $\{\Upsilon _2, \beta _2/H, T_N^{(2)}, v_w^{(2)}  \}$ and the former as $\{\Upsilon _3, \beta _3/H, T_N^{(3)}, v_w^{(3)}  \}$.  
Particularly, we expect the friction within another bubble to be typically less than in the vanilla vacuum, as the fractional change in masses is smaller. There is also a smaller $\Upsilon$ as there is typically a smaller difference between the potentials of the bubbles than between the second bubble and the vanilla vacuum. It is more involved to see the difference in the transition rates, as these depend on the size of the barrier between the transitions. For our benchmarks we find that all transitions are strongly first order. Lastly, we also expect a change in transition temperature between scenarios 2) and 3).

Finally, we want to comment on the collision of bubbles of different vacuum. Such bubbles will not merge, but the bubble corresponding to the deeper vacuum can consume the other. The process can be understood as follows. Near the two bubble walls the field must continuously evolve, in both field space and space-time. It will then be energetically favourable for the bubble wall of the deeper vacuum to advance forward. However, even without collision, the consumption of a bubble may lead a forth contribution to the relic gravitational waves, since the plasma shells of the two different species of bubbles could interact, potentially creating another sound wave contribution. Understanding the potential production of gravitational waves from such a process would require a precise numerical simulation, so we will leave this for future work.

\section{Gravitational waves from multi-step phase transitions}\label{multi-step}
\subsection{Gravitational wave spectrum}
First order cosmological phase transitions proceed via bubble nucleation. Upon collisions of such bubbles, the latent heat will be converted to bulk flow of the plasma, as well as to kinetic energy of the scalar fields. The contributions to the GW spectrum of the latter are well captured by the envelope approximation \cite{Huber:2008hg}, but the contributions from the plasma flow are much harder to capture in a model. Moreover, recent studies indicate \cite{Hindmarsh:2017gnf} that the plasma flow contributions dominate over the scalar field contributions, since the plasma flow continues to source GWs long after the collisions of the bubbles. 

Progress in this area has been largely dominated by large-scale hydrodynamic simulations. Nevertheless, well-motivated simplified models have been developed recently, such as the recent bulk flow model \cite{Jinno:2017fby} and sound shell model \cite{Hindmarsh:2016lnk}. Such models may describe the physics in regimes in where simulations have limitations \cite{Konstandin:2017sat}.

The gravitational wave spectrum produced by a cosmic phase transition can be written in terms of a sum of three contributions - a collision term describing the collision of bubbles, a sound wave term describing the interaction of plasma shells after a collision and finally a turbulence term.
Here, we will adopt a parametrization from \cite{Weir:2017wfa} to capture the essential features of the spectrum as follows:
\be\label{OmegaGWoneT} \Omega_{\rm GW} = \Omega_{\rm env} + \Omega_{\rm sw} + \Omega_{\rm turb} \ee
However, our analysis can straightforwardly adapted when future models become available.
Following \cite{Weir:2017wfa}, the collision term is given by 
\begin{eqnarray}
h^2 \Omega _{\rm col} = 1.67 \times 19^{-5} \left( \frac{0.48 v_w^3}{1+5.3 v_w^2+5v_w^4} \right) \left( \frac{\beta}{H} \right) ^{-2} \kappa _{\rm col}^2  \left( \frac{\Upsilon }{1+\Upsilon } \right)^2 \left( \frac{100}{g_*} \right) ^{1/3} S_{\rm col}(f)
\end{eqnarray}
The efficiency parameter $\kappa _\phi$ is typically vanishingly small, making this contribution sub-dominant. For $v_w$ close to unity one has for the frequency spectrum,
\begin{equation}
   S_{\rm col} = \left[ c_l \left( \frac{f}{f_{\rm col}} \right)^{-3} + (1-c_l-c_h) \left( \frac{f}{f_{\rm col}} \right)^{-1}+ c_h \left( \frac{f}{f_{\rm col}} \right)^{-1} \right] \ . 
\end{equation}
Fitting yields $c_l=0.064$ and $c_h=0.48$ \cite{Weir:2017wfa}, as well as
\begin{equation}
    f _{\rm col} = 1.65 \times 10^{-7} {\rm Hz} \left( \frac{0.35}{1+0.069 v_w+0.69 v_w^4} \right)  \left( \frac{\beta}{H} \right) \left( \frac{T_N}{\rm GeV} \right) \left( \frac{g_*}{100} \right)^{1/6} \ . 
\end{equation}
The sound wave contribution is typically larger. Its power spectrum is
\begin{equation}
h^2\Omega _{\rm sw}    = 8.5 \times 10^{-6} \left( \frac{100}{g_*} \right)^{-1/3} \Gamma ^2 \bar{U}_f^4  \left( \frac{\beta}{H} \right)^{-1}  v_w S_{\rm col}(f) 
\end{equation}
where $\Gamma \sim 4/3$ is the adiabatic index, and $\bar{U}_f^2\sim (3/4) \kappa _f \alpha _T$ is the rms fluid velocity. The efficiency parameter is in this case given by
\begin{equation}
    \kappa _f \sim \frac{\Upsilon }{0.73+0.083 \sqrt{\Upsilon } +\Upsilon}
\end{equation}
and the spectral shape is given by
\begin{equation}
    S_{\rm sw} =  \left( \frac{f}{f_{\rm sw}} \right) ^3 \left( \frac{7}{4+3\left( \frac{f}{f_{\rm sw}}\right) ^2} \right)^{7/2}
\end{equation}
with
\begin{equation}
    f_{\rm sw} = 8.9 \times 10^{-7} {\rm Hz} \frac{1}{v_w} \left( \frac{\beta}{H} \right) \left( \frac{T_N}{{\rm Gev}} \right) \left( \frac{g_* }{100} \right)^{1/6} \ .
\end{equation}
Finally, for the turbulence contribution we have 
\begin{equation}
    h^2 \Omega _{\rm turb}(f) = 3.35 \times 10^{-4} \left( \frac{\beta }{H} \right) ^{-1} \kappa _{\rm turb}^{3/2}  \left( \frac{\Upsilon }{1+\Upsilon} \right)^{3/2} \left( \frac{100}{g_*} \right) ^{1/3} v_w S_{\rm turb} (f) \end{equation}
    where we may take $\kappa  _{\rm turb} \sim 10^{-4}$, making this contribution also sub-dominant. The spectrum is
    \begin{equation}
        S_{\rm turb} = \frac{(f/f_{\rm turb})^3}{\left[ 1+(f/f_{\rm turb})\right]^{11/3}(1+8 \pi f/h*)}
    \end{equation}
    with
    \begin{eqnarray}
    h_* &=& 1.65 \times 10^{-7} {\rm Hz} \left(\frac{T_N}{{\rm GeV}} \right) \left( \frac{g_*}{100} \right)^{1/6} \\
    f_{\rm turb} &=& 27 \times 10^{-7} {\rm Hz} \frac{1}{v_w} \left( \frac{\beta}{H} \right)\left( \frac{T_N}{\rm GeV} \right) \left( \frac{g_*}{100} \right) ^{1/6}
    \end{eqnarray}
The above contributions were estimated in the case of a single phase transition, which completes quickly, so the thermal parameters at the nucleation temperature are to model the gravitational wave spectrum. In our case the thermal parameters vary considerably during the phase transition as can be seen by Fig. \ref{fig:thermalparameters}. This is due to the first phase transition completing over $\Delta T \sim 20$ GeV. As such we expect there to be a large numerical uncertainty in applying these fits to our scenario. This source of uncertainty we leave to future simulations. We will, however estimate the effects of overlapping transitions in the next subsection, based on the spectra of single PTs above.

\subsection{Overlapping phase transitions}
\begin{figure}
    \centering
    \includegraphics[width=0.5\textwidth]{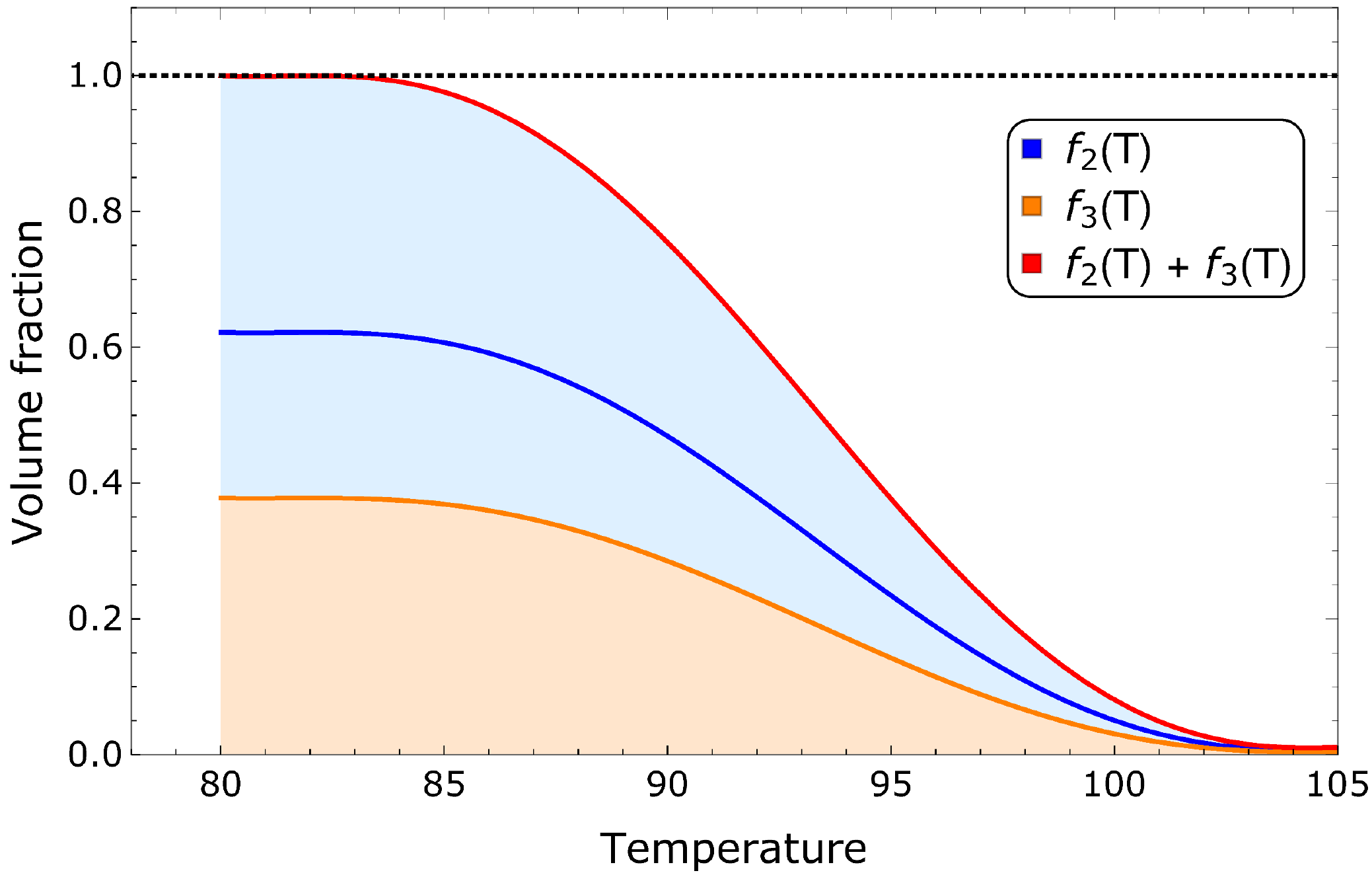}
    \caption{Example of relative weight factors for two nucleation scenarios of the same vacuum.}
    \label{fig:weightfactors}
\end{figure}

In the case of overlapping phase transitions there are different thermal parameters, corresponding to different situations which we cover in this subsection. We will make the simplifying assumption that the thermal parameters are determined by the nucleation scenario of the bubbles.
In reality, the thermal parameters may evolve, which may be modeled in a future numerical analysis. 
Throughout this paper we model the peak amplitude, frequency and spectrum produced by these situations as a weighted sum of the different nucleation situations as follows,
\begin{eqnarray} \label{OmegaGW}
 \Omega _{GW} =  \sum _i w_i\, \Omega _{GW}^{(i)} \left( \Omega _0^{(i)}, f^{(i)},\Upsilon _i , v_w^{(i)}, \frac{\beta ^{(i)}}{H} , T_N ^{(i)}\right) \ .
\end{eqnarray}
Here the index $i$ characterizes the nucleation scenario. 
For example, for the simultaneous PTs to vacua A and B, one may have (conf. Fig. \ref{fig:doublebubble}),
\begin{enumerate}
    \item Vacuum A $(v_A,0)$ in the vanilla vacuum\footnote{We will refer to the vacuum where both fields are at or near the origin as the vanilla vacuum: $(0,0)$.} $(i=1)$;
    \item Vacuum B $(v_A,v_B)$ in the vanilla vacuum $(i=2)$;
    \item Vacuum B $(v_A,v_B)$ in vacuum A $(i=3)$.
\end{enumerate}
Each of these situations will in principle be described by different parameters $(v_w^{(i)}, \Upsilon_i, \beta^{(i)})$.
Then, $w_i(t)$ characterizes the relative importance of the nucleation scenarios, which we will estimate in the following way. 
First, we note that if there are different nucleation scenarios of the same vacuum, the definition of $T_f$ (Eq. \eqref{Tf}) should change, to reflect that the PT finishes when the vacuum has filled all of space-time, 
\be \label{newTf1} \mathop{\mathlarger{\sum}}\limits_{j \in \text{vacuum$(i)$}} f_j\left(\tilde{T}_f^{(i)}\right)=1  \,\,\,\,\,\,\,\,\,\,\,\, \text{PT completes} \ee
where $f_i(T)$ is evaluated at $v_w^{(i)}, \Upsilon_i, \beta^{(i)}$, corresponding to the nucleation scenario. Explicitly, in the above example, this means that transitions 2. and 3. both finish at $\tilde{T}^{(2)}_f=\tilde{T}^{(3)}_f$, when $$f_2\left(\tilde{T}^{(2)}_f\right) + f_3\left(\tilde{T}^{(2)}_f\right) =1.$$ For a schematic representations, see Fig. \ref{fig:weightfactors}.

It is important to note that for simultaneous transitions, the earlier PT may not reach completion, as the consecutive PT takes over before $f_i(T)=1$ is reached. In the above example, this applies to the first transition. In such cases Eq. \eqref{Tf} also breaks down. Instead, we may estimate the final temperature $\tilde{T}_f$ by
\be \label{newTf2}\mathop{\mathlarger{\sum}}\limits_{j \in \text{vacuum$(i)$}} f_j\left(\tilde{T}_f^{(i)}\right) +  \mathop{\mathlarger{\sum}}\limits_{k \not\in\text{vacuum$(i)$}} f_k\left(\tilde{T}_f^{(i)}\right) =1 \,\,\,\,\,\,\,\,\,\,\,\, \text{PT does not complete}\ee
such that the earlier PT is cut-off when the later PT has caught up.   

Thus, summarizing \eqref{newTf1} and \eqref{newTf2}, the final temperatures of simultaneous PTs can be estimated by,
\be \label{newTf} \mathop{\mathlarger{\sum}}\limits_{j \in \text{vacuum$(i)$}} f_j\left(\tilde{T}_f^{(i)}\right) = \begin{cases} 1 &\text{PT completes} \\ 1- \mathop{\mathlarger{\sum}}\limits_{k \not\in\text{vacuum$(i)$}} f_k\left(\tilde{T}_f^{(i)}\right) &\text{PT does not complete}  \end{cases}\ee

Then, the weight factor can be approximated by,
\be\label{omegai} w_i  =\frac{ \mathop{\mathlarger{\int}}^{\tilde{T}^{(i)}_f}_{T^{(i)}_N}  f_i(T) dT}{ \mathop{\mathlarger{\int}}_{\tilde{T}^{(i)}_N}^{\tilde{T}^{(i)}_f} \left[ \mathop{\mathlarger{\sum}}\limits_{j \in \text{vacuum$(i)$}}f_j(T)\right] dT} \,\,\,\,\,\,\,\,\, \text{PT completes} \ee
for PT which complete. It is seen that for consecutive phase transitions, which are separated in time, there is only a single nucleation scenario for each vacuum, such that \eqref{omegai} gives $w_i = 1, \, \,\forall i$ as expected. The spectrum \eqref{OmegaGW} is in this case a simple sum of the individual spectra. In the case that there are different nucleation scenarios of a vacuum, the $w_i$ sum to one. In the above example, we would have $w_2 + w_3 =1$.
It is seen that any deviations in the spectrum will feature most strongly for $w_{i} \approx w_{j}, \,\, j \in\text{vacuum}(i)$.

However, if a PT does not complete, because the following PT takes over, the GW spectrum resulting from this PT will only be a fraction of the spectrum which would have resulted, if the PT had been allowed to finish. Therefore, the
$w_i$ corresponding to such a vacuum can not sum to one. Here we will estimate it by,
\be\label{omegaiN} w_i  =\frac{ \mathop{\mathlarger{\int}}^{\tilde{T}^{(i)}_f}_{T^{(i)}_N}  f_i(T) dT}{ \mathop{\mathlarger{\int}}_{T^{(i)}_N}^{T^{(i)}_f} f_i(T) dT} \,\,\,\,\,\,\,\,\, \text{PT does not complete} \ee
where $T_f$ corresponds to the end of the vanilla transitions \eqref{Tf} and $\tilde{T}_f$ Eq. \eqref{newTf}
This insight is an important result of this analysis. It implies that the GW spectrum from simultaneous PT can not be calculated as the sum of consecutive PT. 

We note here that we have not taken into account that bubbles of the same vacuum, from different nucleation scenarios, may collide, such that $f^{(i)}(T)$ may overlap. Also, \eqref{newTf2} gives an upper bound of the final temperature of a PT which does not complete.
As a consequence, the above Eq. \eqref{newTf} may overestimate the final temperature $T_f$. Eq. \eqref{omegai} may therefore be considered a conservative estimate of the effect of the simultaneous PT.

\subsection{Effects in a toy model}
Here we will demonstrate the effects described in the previous section in a toy model. Our toy model will be a hypothetical scenario with two singlet PTs, and is described by the example of the previous sections, and Fig. \ref{fig:doublebubble}. 

\begin{figure}[H]
    \centering
    \includegraphics[width=.45\textwidth]{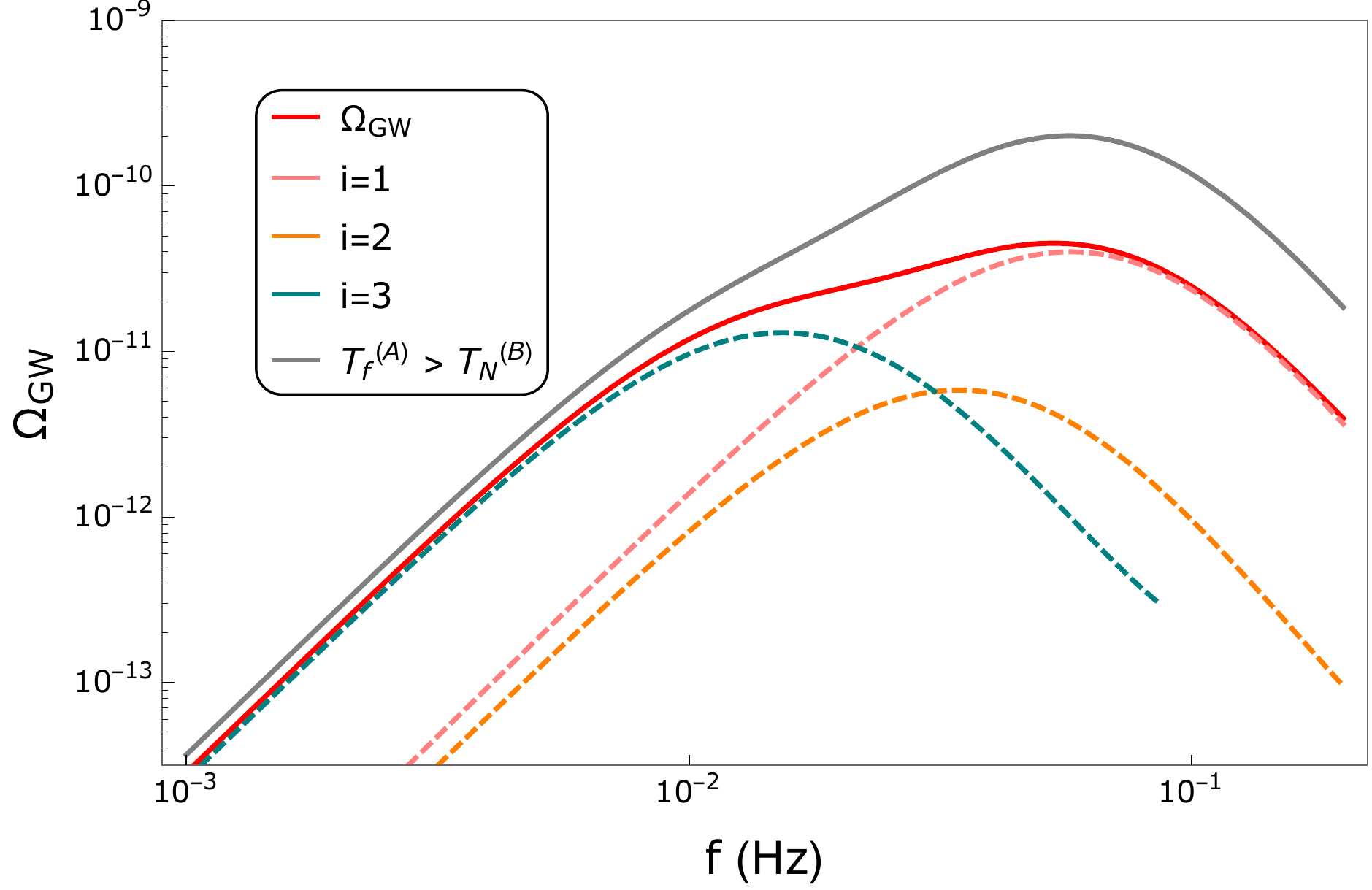}
    \includegraphics[width=.45\textwidth]{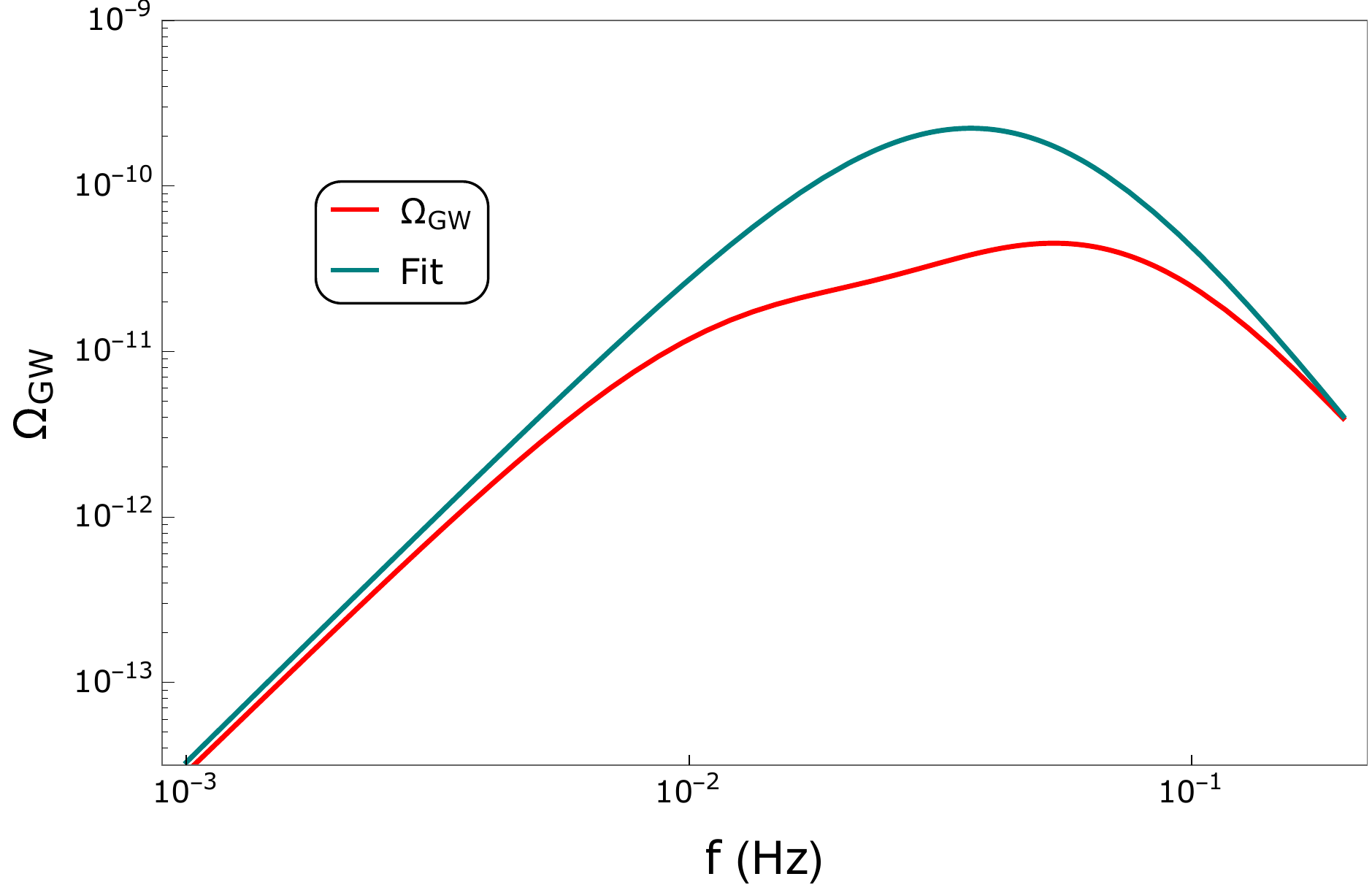} \\
    \includegraphics[width=.45\textwidth]{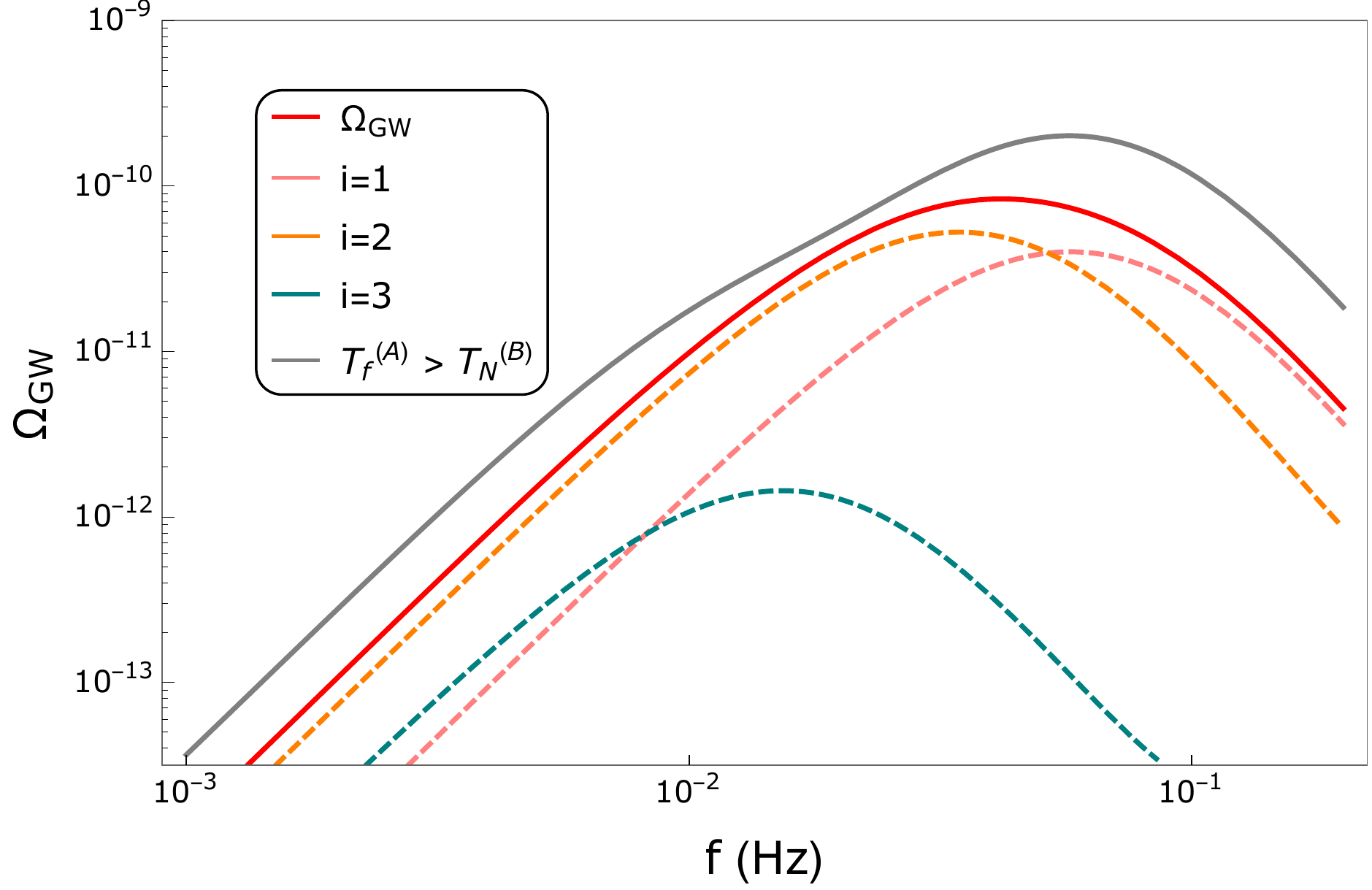}
    \includegraphics[width=.45\textwidth]{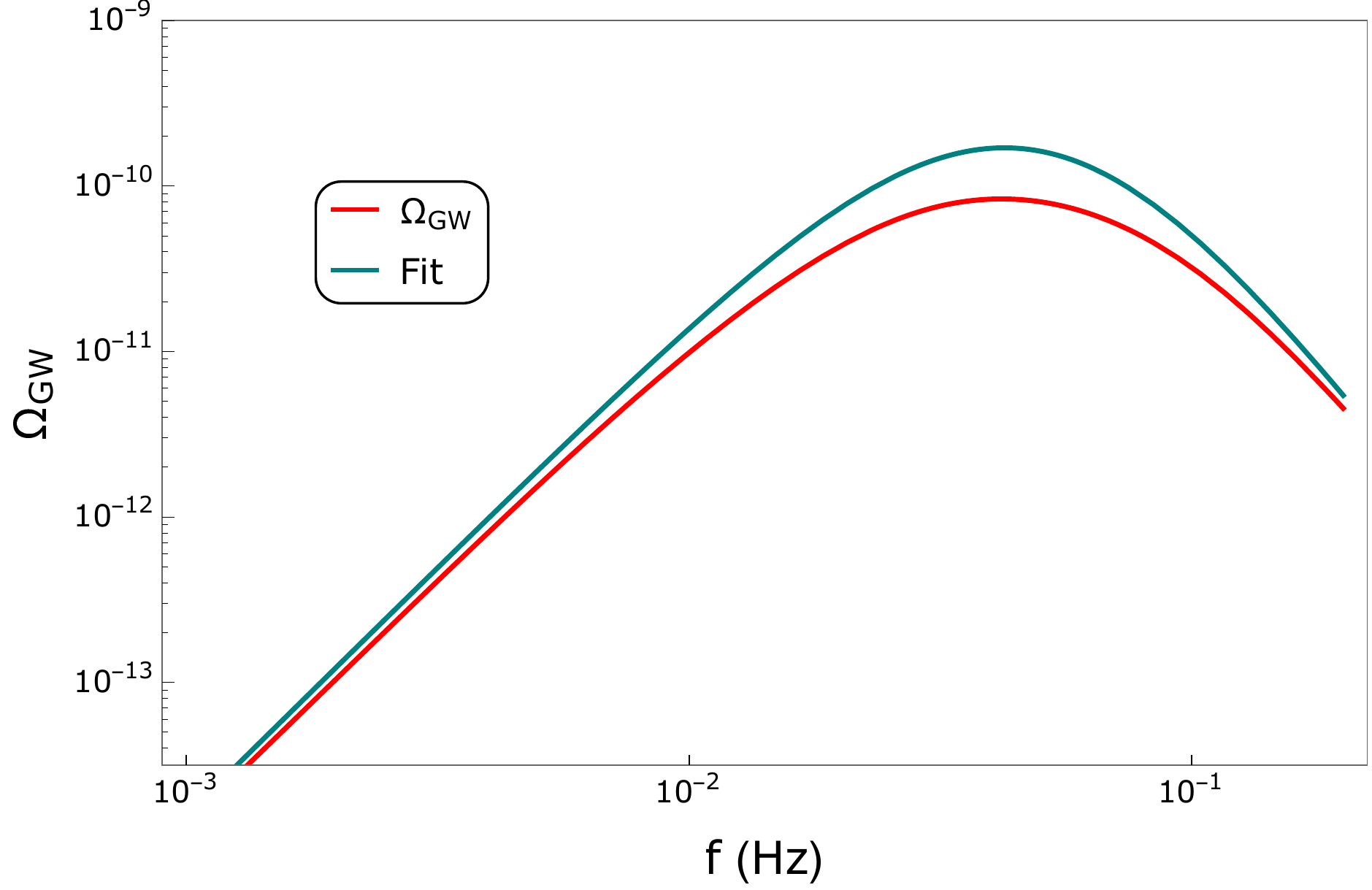}
    \caption{Left panels: the spectrum due to two simultaneous PTs, given by the red line, leads to a different spectrum than consecutive transitions. This can be seen by comparison with the gray line, which is predicted by the same thermal parameters, and $w_1 = w_3 = 1$, $w_2=0$. In the upper example we chose $w_2=0.1=1-w_3$. In the lower example we chose $w_2=0.9=1-w_3$. In both examples, $w_1 = 0.2$. Right panels: the spectrum from simultaneous PTs can not be fitted to the spectrum than from a single PT. Shown are fits to $\Omega_{GW}$ in Eq. \eqref{OmegaGWoneT} with $\Upsilon = 0.22$, $v_w=0.65$, $\beta/H=124$, $T_N=134$ GeV (upper panel), and $\Upsilon = 0.21$, $v_w=0.65$, $\beta/H=142$, $T_N=136$ GeV (lower panel). }
    \label{fig:effects}
\end{figure}

The first and most prominent effect of overlapping transitions, is that the first transition may not complete. Therefore, the weight parameter $w_1 < 1$, and the overall spectrum is expected to be weaker. This effect may also cause the double peak of the spectrum to disappear, as shown in the lower left panel of Fig. \ref{fig:effects}. However, as is shown in the lower right panel of Fig. \ref{fig:effects}, this spectrum can still not be mimicked by the spectrum from a single PT. 

Whether the spectrum has a double peak or a ``shoulder", depends primarily on the difference in nucleation temperature $T_N$ between the different scenarios. It is expected that $T_N^{(2)}>T_N^{(3)}$, i.e., the nucleation temperature inside a bubble of vacuum A is larger than in the vanilla vacuum. Therefore, the double peak is expected to appear for $T_N^{(3)}\ll T_N^{(1)}$ and $w_3 \rightarrow 1$, which coincides with the limit of consecutive PT. This is shown in the lower panels of Fig. \ref{fig:effects}. 

\begin{figure}[t]
    \centering
    \includegraphics[width=.55\textwidth]{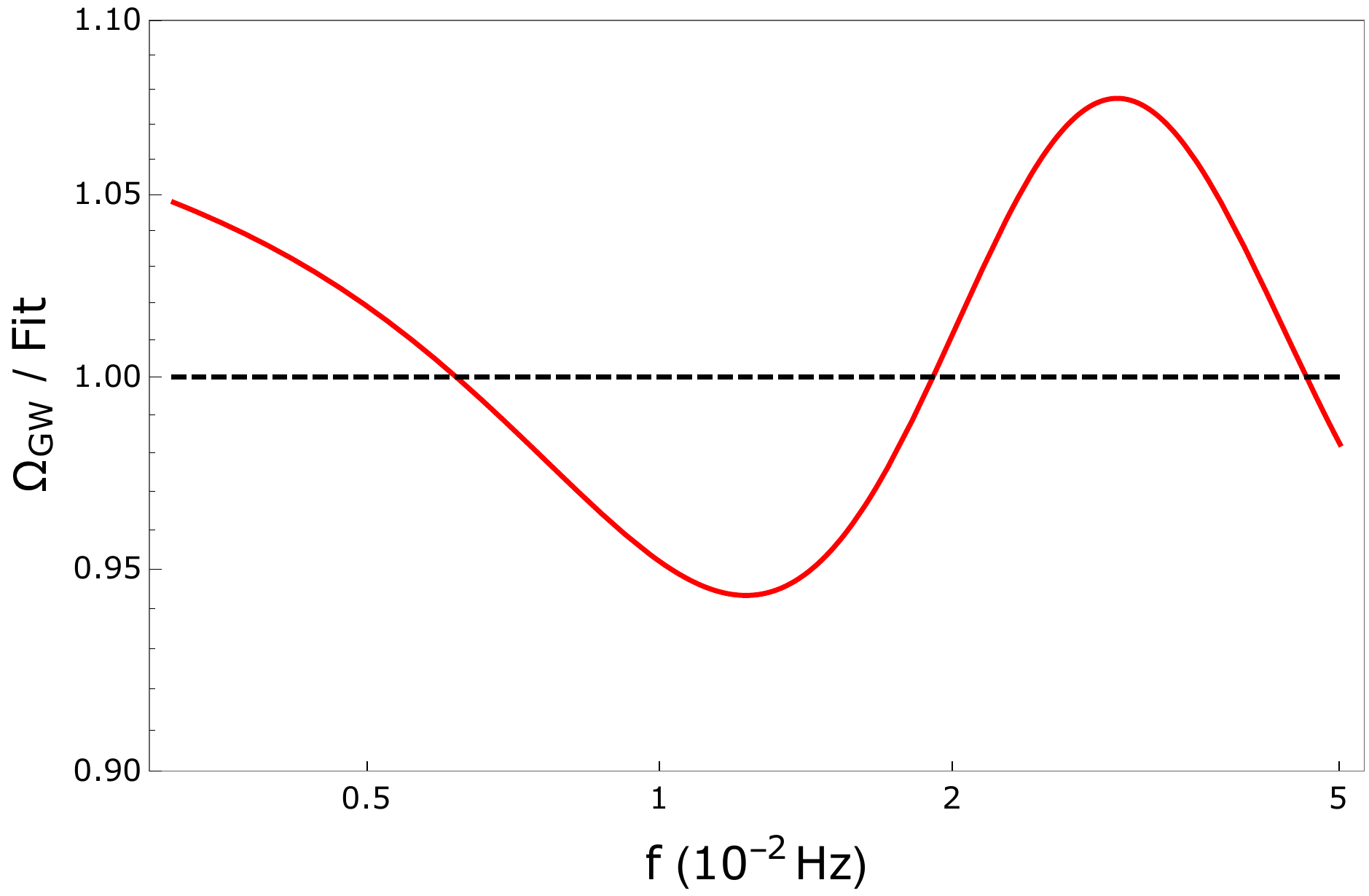}
    \caption{For $w_2=w_3 =0.5$, the spectrum from simultaneous PTs can not be fitted to the spectrum of two single PTs. The comparison is to a fit of $\Omega_{GW}= \Omega_{GW}^{(A)} + \Omega_{GW}^{(B)}$ with the parameters $\Upsilon = 0.094$, $v_w=0.65$, $\beta/H=77.3$, $T_N=115$ GeV (transition A), and $\Upsilon = 0.16$, $v_w=0.65$, $\beta/H=179$, $T_N=133$ GeV (transition B). The frequency band fitted and shown has been chosen such that $\Omega_{GW}>10^{-12}$.}
    \label{fig:doublefit}
\end{figure}

One may wonder whether whether the spectrum can be fitted to two consecutive PT. This depends on the relative importance of the weight factors $w_2$ and $w_3$. For $w_2 \sim w_3$, it is generically not possible to mimic the effect of simultaneous PTs by two consecutive PT, as demonstrated in Fig.\ref{fig:doublefit}. 
We note here that the figure shows an idealized situation with an exaggerated difference in nucleation temperatures, to visualize the effect. We will turn to a more realistic model in the following section. 

\section{Standard Model plus singlet example}\label{singletexample}
The Singlet extension to the Standard Model is not the ideal scenario to demonstrate the effect of an overlapping phase transition visible at LISA as the nucleation temperature tends to be a little low rendering the signal at the edge of what LISA can detect. Nonetheless it is useful as an example to demonstrate how this effect can occur in the simplest extension to the Standard Model as well as a scenario that is desirable for electroweak baryogenesis \cite{Profumo:2014opa}. The effective potential can be broken up as follows 
\begin{equation}
    {\cal V}= V_{\rm SM} + V_{\rm portal} + V_{\rm singlet}
\end{equation}
where 
\begin{eqnarray}
V_{\rm SM} &=& \mu ^2 H^2 + \frac{\lambda }{4} H^4 \\
V_{\rm portal } &=& \frac{\kappa _1}{2} S H^2 + \frac{\kappa _2}{2} S^2 H^2 \\
V_{\rm Singlet} &=& M^2 S^2 + \kappa S^3 +\lambda _s S^4 \ .
\end{eqnarray}
In our particular choice of the Lagrangian there are terms that explicitly break any $Z_2$ symmetry which avoids the problem of cosmologically catastrophic domain walls (see \cite{Mazumdar:2015dwd} for a discussion of various solutions to domain wall problems). 
The two parameters in the Standard Model potential are fixed by the mass of the Higgs and the W boson (which determines the Standard Model VEV) respectively. In the Standard Model one has at the electroweak scale $O(100)$ GeV, $\mu ^2 \sim 3848$ GeV and $\lambda \sim 0.25$. When the singlet acquires a vacuum expectation value, one needs to correct the values of the Higgs sector parameters such that the mass of the Higgs and W boson coincide with their measured values.
 For the mass range we are interested in $|\sin \theta| \lesssim 0.2$ \cite{Chalons:2016jeu,Ilnicka:2018def} although future colliders are expected to improved this bound \cite{Chang:2018pjp,Kotwal:2016tex}. In principle, a lower bound on the mixing angle can be found from BBN, since long lived particles may interfere with light element production \cite{Fradette:2017sdd}. However, for the parameter space we consider here, the portal couplings have no effect on BBN, but are instead probed by colliders. The range of portal couplings needed to catalyse a strongly first order electroweak phase transition typically requires $|\kappa 1|$ to be negative with the approximate range $-500 \lesssim \kappa 1 \lesssim -100 $ and $\kappa _2 \sim O(1)$. Further more ref. \cite{Profumo:2014opa} found that the portal couplings are anti-correlated for phenomelogically viable parameter sets that result in a strongly first order electroweak phase transition. This anti-correlation suppresses the off diagonal elements in the mass matrix causing the mixing angle to be small.\par 
Finally for the singlet part of the potential it is convenient to reparametrize the potential such that the inputs of the potential are the scale of the potential, the local minimum and a parameter which controls the size of the barrier between the origin and the singlet minimum. Such a parametrization was given in ref. \cite{Akula:2016gpl}
\begin{equation}
    V_{\rm Singlet} = \Lambda ^4 \left[\frac{3-4 \alpha }{2} \left( \frac{s^2}{v_s^2} \right)-\left( \frac{s^3}{v_s^3} \right)+\alpha \left( \frac{s^4}{v_s^4} \right) \right] \ .
\end{equation}
Where scale of the potential is given by $\Lambda$, the singlet vev is denoted $v_s$ and $\alpha \in [0.5,0.75]$ parametrizes the size and location of the tree level barrier between the true ($v_s$) and false vacuum ($0$). Here $\alpha =0.5$ corresponds to degenerate minima and $\alpha = 0.75$ corresponds to the case where there is no barrier. The critical temperature and the mass of the singlet are both approximately the scale of the potential. That is $\Lambda \sim M_s \sim T_N$ up to an order one factor. Since we want the transition to overlap with the electroweak phase transition we require $\Lambda \sim m_H$. \par The nucleation temperature, $T_N$, is lower than the critical temperature by an amount
\begin{equation}
    T_N = (1-\epsilon ) T_C
\end{equation}
where a large epsilon corresponds to the case where there is a lot of supercooling. The amount of supercooling is largest for smaller $\alpha$ and larger $v_S$. To ensure the two phase transitions are close together we require $v_S \sim v_{\rm SM}$. Finally the strength of the phase transition is parametrized by the order parameter $\phi _C/T_C$ and is controlled by $\alpha$ and $v_S$ where a larger $v_S$ and a smaller $\alpha$ correspond to a stronger phase transition.

\begin{table}[b]
    \centering
    \begin{tabular}{c|c|c|c|c|c}
        & $\Upsilon _i$ & $v_w$ & $T_N^i$ & $\beta /H$  \\ \hline $i=1$  & $0.22$ & $0.69$ & $120$ &  $29.0$   \\
        $i=2$  & $0.20$& $ $ & $72.0$ & $13.8$  \\
        $i=3$ & $0.076$ & $ $ & $61.3$ & $161.5$ \\ \hline 
        $i=1$  & $0.14$ & $0.84$ & $116$ & $33.2$    \\ 
        $i=2$   & $0.10$ & $ $ & $70.0$ & $40.8$    \\ 
        $i=3$  & $0.075$ & $ $ & $65.0$ & $128$    \\ \hline 
        $i=1$  & $0.29$ & $0.58$ & $106$ & $42.7$    \\ 
        $i=2$   & $0.10$ & $ $ & $70.0$ & $4.10$    \\ 
        $i=3$  & $0.077$ & $ $ & $57.0$ & $183$  
    \end{tabular}
    \caption{Inputs to relic gravitational wave spectrums for all four SM + singlet benchmarks}
    \label{tab:benchmwarksGW}
\end{table}
\begin{table}[t]
    \centering
    \begin{tabular}{c|c|c|c|c|c}
         $\kappa _1$ & $\kappa _2$ & $\Lambda $ & $\alpha $ & $v_S$ & $\sin \theta $  \\ \hline 
        $-295$ & $0.850$ & $120$ & $0.61$ & $299$ & $0.18$ \\ \hline 
        $-300$ & $0.900$ & $109$ & $0.62$ & $300$ & $0.11$ \\ \hline 
        $-390$ & $1.35$ & $120$ & $0.59$ & $249$ & $0.074$ 
    \end{tabular}
    \caption{Lagrangian parameters for the four benchmarks for SM+S as well as the mixing angle $\sin \theta$. Higgs sector parameters are determined by fixing the mass of the Higgs and W boson. All dimensionful parameters are given in GeV.}
    \label{tab:benchmarksL}
\end{table}

\subsection{High temperature corrections}
For simplicity we assume that the mixing is sufficiently small (for our benchmarks $\sin \theta <0.1$) and that we can use the high temperature expansion in the Landau gauge\footnote{For a discussion on gauge invariance we refer the reader to \cite{Patel:2011th}.} such that the temperature corrections to the effective potential can be approximated in the high temperature expansion and the limit of no mixing as
\begin{eqnarray}
V_{\rm SM} &\to & (\mu ^2 +c_h T^2) h^2 -E_hT h^3+ \frac{\lambda }{4} H^4 \\
V_{\rm portal } &\to & \kappa _1 S H^2 + \frac{\kappa _2}{2} S^2 H^2 \\
V_{\rm Singlet} &\to & \Lambda ^4 \left[\left( \frac{3-4 \alpha }{2} +c_s \frac{T^2v_s^2}{\Lambda^4} \right) \left( \frac{s^2}{v_s^2} \right)-\left(1+\frac{E_s T v_s^3}{\Lambda ^4} \right)\left( \frac{s^3}{v_s^3} \right)+\alpha \left( \frac{s^4}{v_s^4} \right) \right] \ .
\end{eqnarray}
In he above we have replaced the quantum fields with the classical fields and defined
\begin{eqnarray}
c_h &=& \frac{1}{32} \left(g_1^2 +3 g_2^2+4y_t^2+4 \lambda \right) \\ 
E_h &=& \frac{3}{96 \pi} \left(2 g_2 ^3 +(g_1^2+g_2^2)^{3/2}  \right) \\
c_s &=& \frac{\alpha}{2} \frac{\Lambda ^4}{ v^4_s} \\
E_s &=& \frac{2\sqrt{3}}{\pi} \frac{\Lambda^6}{v_s^6} \alpha ^{3/2} \ .
\end{eqnarray}
In this case the system can be solved semi-analytically, as we demonstrate in the next subsection. Since this work is a proof of principle we leave precise calculations of these phase transitions to future work and we expect our benchmark to approximate the parameters which describe our scenario.
\subsection{Semi-analytic treatment of transitions}
There are three possible transitions that we are interested in for the case where the singlet has the higher nucleation temperature. They are 
\begin{eqnarray}
(0,0) &\to & (0,v_s) \nonumber \\
(0,0) &\to & (v_h,v_s^\prime ) \nonumber \\
(0,v_s) & \to & (v_h,v_s^\prime ) \ .
\end{eqnarray}
The first transition is straightforward to deal with analytically. 
The change in the scale of the potential, $\Lambda$, the minima $v_s$ and the parameter $\alpha $ can be determined by simultaneously solving
\begin{eqnarray}
\frac{\Lambda ^4}{v_s^2} \left(\frac{3-4 \alpha }{2} + \frac{c_s T^2v_s^2}{\Lambda ^4} \right) &=& \frac{\Lambda ^4(T)}{v_s^2(T)} \left( \frac{3-4\alpha (T)}{2} \right) \\ 
\frac{\Lambda ^4}{v_s^3}\left( 1 + \frac{E_s T v_s^3}{\Lambda ^4} \right) &=& \frac{\Lambda ^4(T)}{v_s^3(T)} \\
\frac{\alpha  \Lambda ^4 }{v_s^4} &=& \frac{\alpha (T) \Lambda ^4(T)}{v_s^4(T)}
\end{eqnarray}
where the right hand side contain the temperature dependent versions of the parameters on the left hand side. 
Note the critical temperature is defined as $\alpha (T_C)=0.5$.  Under these definitions the singlet potential becomes
\begin{equation}
    V_{\rm Singlet} (T) = \Lambda (T) ^4 \left[\frac{3-4 \alpha (T) }{2} \left( \frac{s^2}{v_s^2(T)} \right)-\left( \frac{s^3}{v_s^3(T)} \right)+\alpha (T) \left( \frac{s^4}{v_s^4(T)} \right) \right] \ .
\end{equation}
To a reasonable approximation one can calculate the bounce as straightline transition in field space from the true to the false vacuum. In this case the bounce profile can be calculated to within a fraction of a percent to be 
\begin{eqnarray}
    s(r,T) &=& \frac{s_0[\alpha(T)]}{2} \left\{1-\tanh \left[ \frac{r \Lambda ^2(T)/v_s(T) -\delta [\alpha(T)]}{L_w[\alpha(T)]} \right] \right\}\nonumber \\ &-& \frac{1}{2} \left| \frac{s_0[\alpha (T)]}{L_w[\alpha(T)]}   {\rm Sech} ^2 \left\{ \frac{\delta [\alpha(T)] }{L_w[\alpha [T]} \right\} \right| e^{-r \Lambda^2(T)/v_s(T)}
\end{eqnarray}
note that $L_w$ and $\delta$ are dimensionless in the above equation. Curve fitting for the above gives 
\begin{eqnarray}
s_0[\alpha (T)] &=& a\left(1- {\rm Tanh} \left[ \frac{\alpha (T) - b}{c} \right]\right)+d \\
L_w[\alpha (T)] &=& a+ b \frac{|\alpha (T)-0.5|^c}{|\alpha (T)-0.75|^d} \\ 
\delta[\alpha (T)] &=& \sum _i \frac{a_i}{|\alpha (T) - b_i|^{c_i}}
\end{eqnarray}
The effective action is then given by
\begin{equation}
    \frac{S_E}{T} = \frac{v_s(T)}{T} \frac{v_s^2(T)}{\Lambda ^2(T)}10^{g[\alpha (T) ]} 
\end{equation}
\begin{table}[t]
    \centering
    \begin{tabular}{c|c|c|c}
        $s_0$ & $\delta $ & $L_w$& g   \\ \hline
        $a=1.50$ & $a_i=\{0.301,2.67,0.711\}$ & $a=1.97$ & $a=-71.06$ \\
        $b=0.718$ & $b_i=\{0.749,1.53,1.69\}$ & $b=6.70$& $b=71.62$ \\  
        $c=0.0854$ & $c_i=\{ 0.558,-77.9,-15.7 \}$ & $c=2.10$& $c=0.008805$ \\  
        $d=-0.479$ &  & $d=0.58$& $d=0.009263$ \\  
    \end{tabular}
    \caption{List of parameters involved in curve fitting the bounce solution (need to fix sig figs later)}
    \label{tab:my_label}
\end{table}
where 
\bea
    g[\alpha (T)] &=&a+ b \frac{|\alpha (T)-0.75|^c}{|\alpha (T)-0.5|^d}
\eea
\begin{figure}
    \centering
    \includegraphics[width=0.4\textwidth]{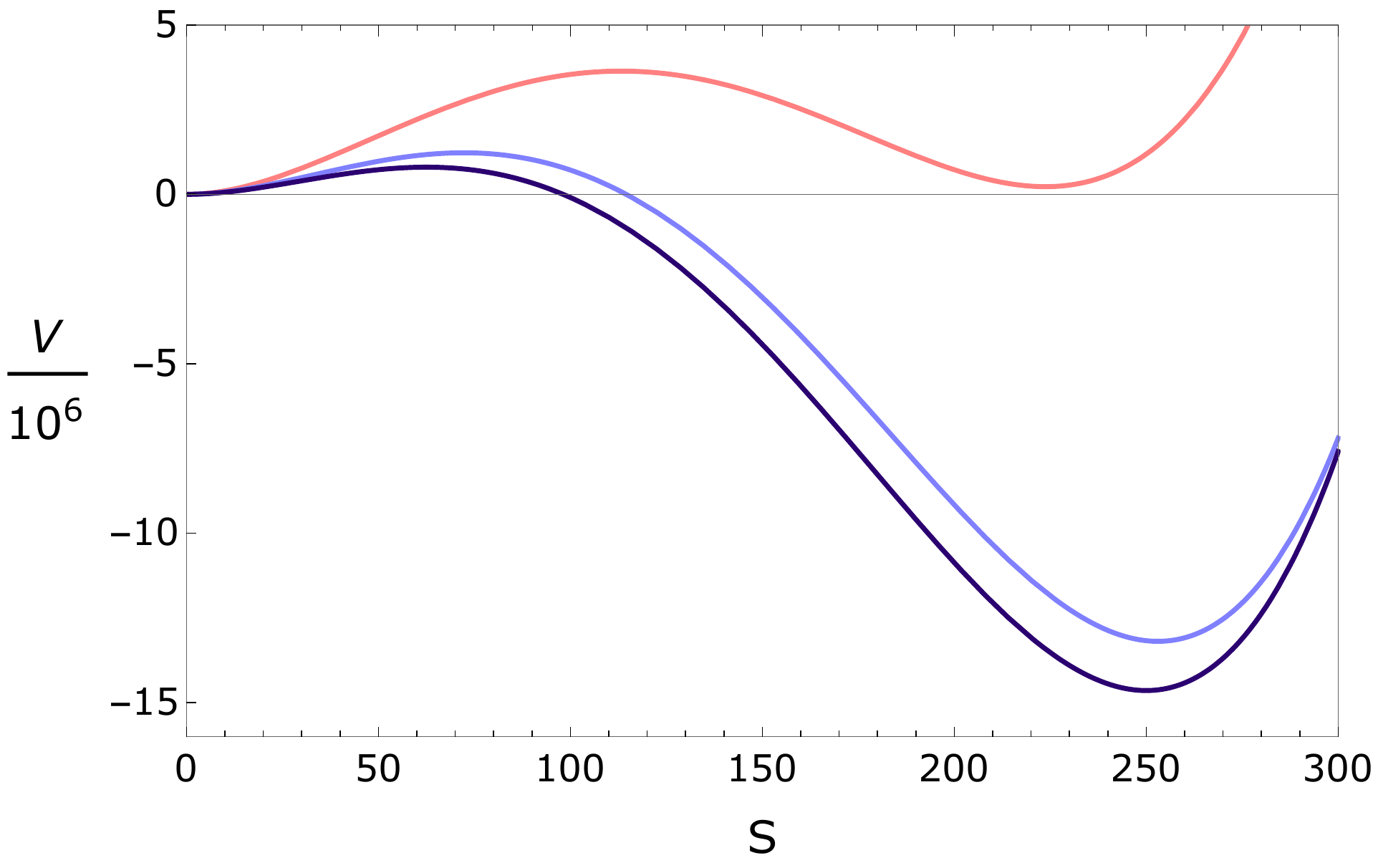}
    \includegraphics[width=0.4\textwidth]{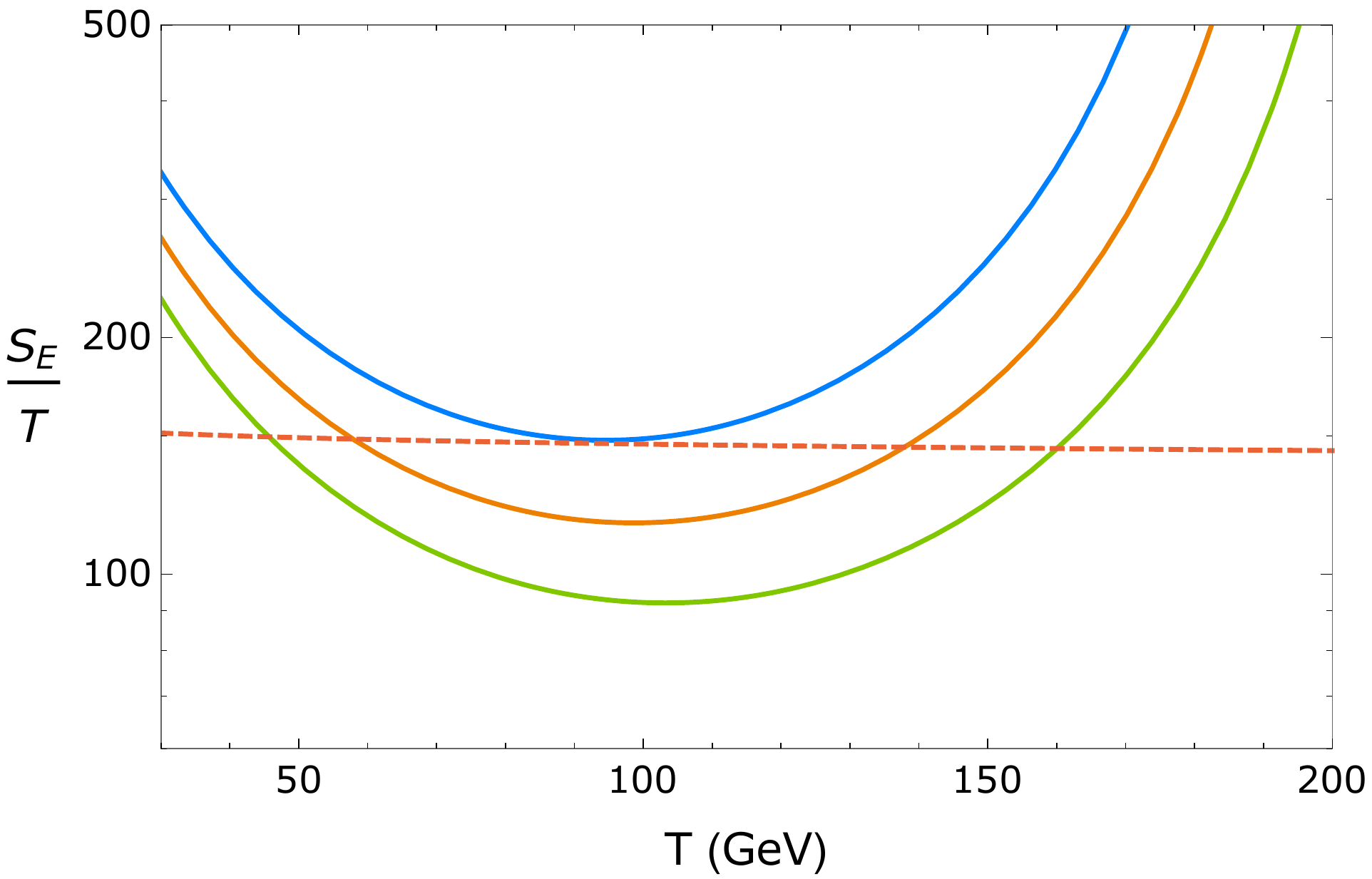} \\
    \includegraphics[width=0.4\textwidth]{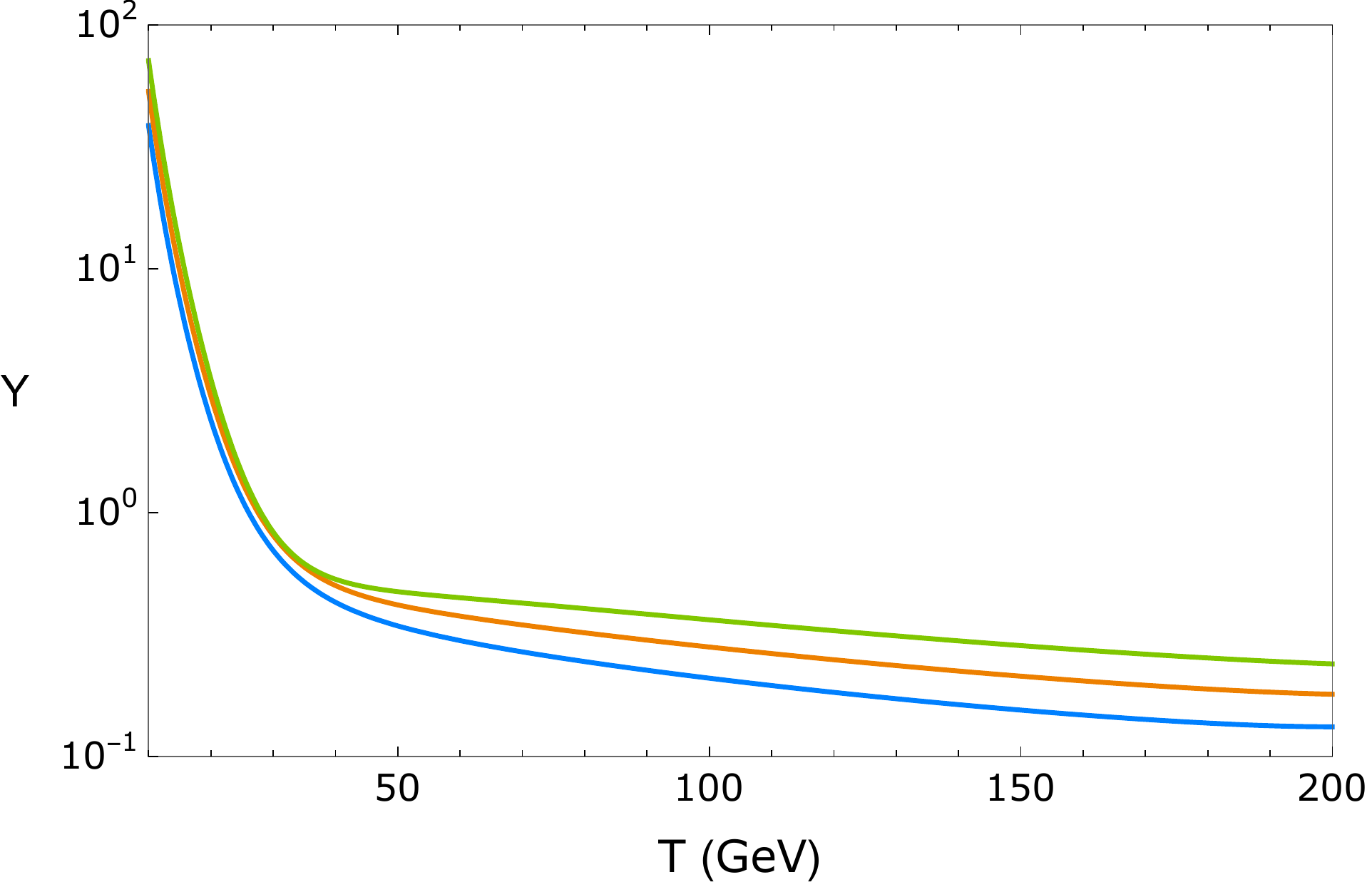}
    \includegraphics[width=0.4\textwidth]{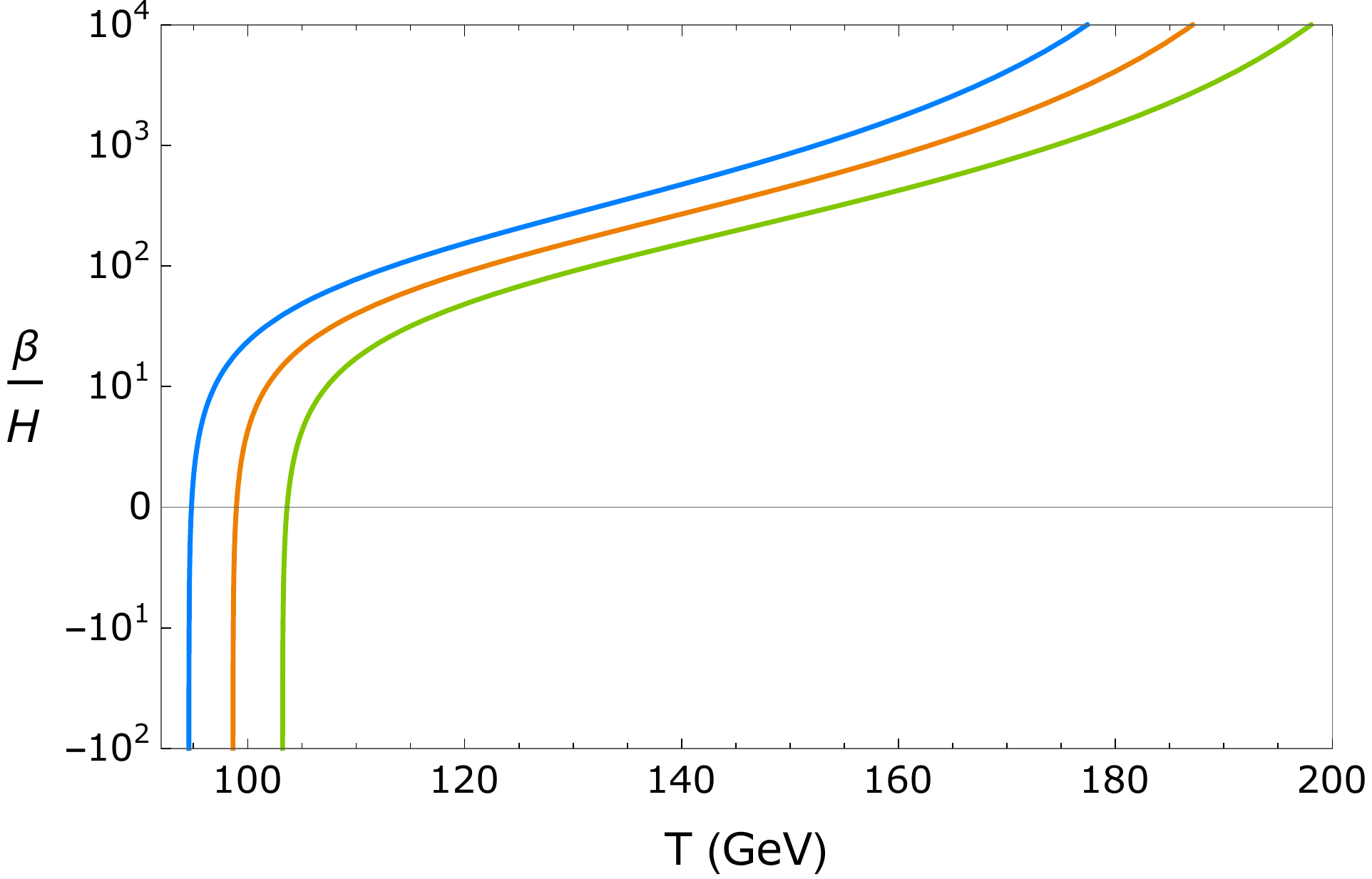}
    \caption{Top left: Evolution of the singlet potential with temperature (top left). The pink curve denotes the potential at the critical temperature, dark blue at the nucleation temperature and light blue at zero temperature. Top right: the evolution of the effective action, bottom left: $\Upsilon$, bottom right: the speed of the phase transition. All with temperature for $v_s=260,\alpha = 0.6$ and $\Lambda=\{115,125,135\}$ for the green, orange and blue lines respectively.}
    \label{fig:thermalparameters}
\end{figure}
The nucleation temperature is of course given when equation \eqref{eqn:TN} is satisfied. 
Let us now turn our attention to the second phase transition where the vanilla vacuum breaks electroweak symmetry and the singlet acquires a vacuum expectation value simultaneously. In this case we can perform the following rotation
\begin{eqnarray}
h &=& \frac{v _h(T)\phi +v_s(T)\theta }{\sqrt{v_s^2(T)+v_h^2(T)}} \\
s &=& \frac{-v_h(T)\phi +v_s(T)\theta }{\sqrt{v_s^2(T)+v_h^2(T)}} 
\end{eqnarray}
where $(v_h(T),v_s(T))$ is the true vacuum at temperature $T$. Setting $\phi = 0$ the potential then looks like
\begin{eqnarray}
\tilde{V} &=&  M_\theta (T) \theta ^2 +\kappa _\theta \theta ^3 + \lambda _\theta \theta ^4 \\
&=&\Lambda _\theta (T) ^4 \left[\frac{3-4 \alpha _\theta (T) }{2} \left( \frac{\theta ^2}{v_\theta ^2(T)} \right)-\left( \frac{\theta^3}{v_\theta ^3(T)} \right)+\alpha _\theta (T) \left( \frac{\theta ^4}{v_\theta^4(T)} \right) \right] \ .
\end{eqnarray}
We can then use the same semi-analytic techniques to calculate how $S_E/T$ evolves with temperature. 
Finally let us turn our attention to the last transition where a singlet bubble tunnels to a bubble where the singlet has a new minimum and electroweak symmetry is broken simultaneously. Note that this transition may be second order even if the other two are first order. For all of our benchmarks we find it to also be first order. We can treat this transition by making the shift $s^\prime = s+ v_s(T)$ after which we perform similar rotations to before
\begin{eqnarray}
h &=& \frac{v _h(T)\phi ^\prime +v_s^\prime(T)\theta ^\prime }{\sqrt{[v_s^\prime]^2(T)+v_h^2(T)}} \\
s^\prime &=& \frac{-v_h(T)\phi^\prime +v_s^\prime(T)\theta ^\prime }{\sqrt{[v_s^\prime]^2(T)+v_h^2(T)}} 
\end{eqnarray}
where $(v_h(T),v_s^\prime(T))$ is the true vacuum at temperature $T$. Setting $\phi ^\prime = 0$ and dropping primes to avoid clutter we once again have
\begin{eqnarray}
\hat{V} &=&  M_\theta (T) \theta ^2 +\kappa _\theta \theta ^3 + \lambda _\theta \theta ^4 \\
&=&\Lambda _\theta (T) ^4 \left[\frac{3-4 \alpha _\theta (T) }{2} \left( \frac{\theta ^2}{v_\theta ^2(T)} \right)-\left( \frac{\theta^3}{v_\theta ^3(T)} \right)+\alpha _\theta (T) \left( \frac{\theta ^4}{v_\theta^4(T)} \right) \right] \ .
\end{eqnarray}

\subsection{Gravitational wave spectrum}
For the Standard Model and singlet scenario, the evolution of the effective potential is shown in Fig. \ref{fig:thermalparameters} (top left). The evolution of the effective action for the singlet transition is given in the top right panel of Fig. \ref{fig:thermalparameters}. Note that there is a minimum in the effective action. To allow for bubble nucleation, the effective action needs to evolve below the critical value (dotted red line), to allow for bubble nucleation. The PT lasts a long interval $\Delta T$ for a more shallow curve, which only dips a small amount below the critical value. Our benchmarks in Table \ref{tab:benchmwarksGW} correspond to such scenarios. 

The other panels of Fig. \ref{fig:thermalparameters} show the evolution of $\Upsilon$ and $\beta/H$ with temperature. Note in particular that $\beta/H$ and $\Upsilon$ change quite quickly during the phase transition. This suggests that there may be some theoretical uncertainty in the gravitational wave spectrum that will be produced, a question which we leave for future numerical simulations. Moreover, $\beta/H$ is fairly small - as you would expect for a phase transition that last a large interval $\Delta T$. \par 
The first  benchmark corresponds to the same singlet parameters as the volume fraction of the singlet bubbles we have shown in Fig. \ref{benchmark1}. In this case, the electroweak phase transition occurs fairly quickly when the singlet bubbles occupy a volume fraction of $\sim 0.5$. The other benchmarks correspond to similar scenarios. \par
\begin{figure}[t]
    \centering
    \includegraphics[width=.45\textwidth]{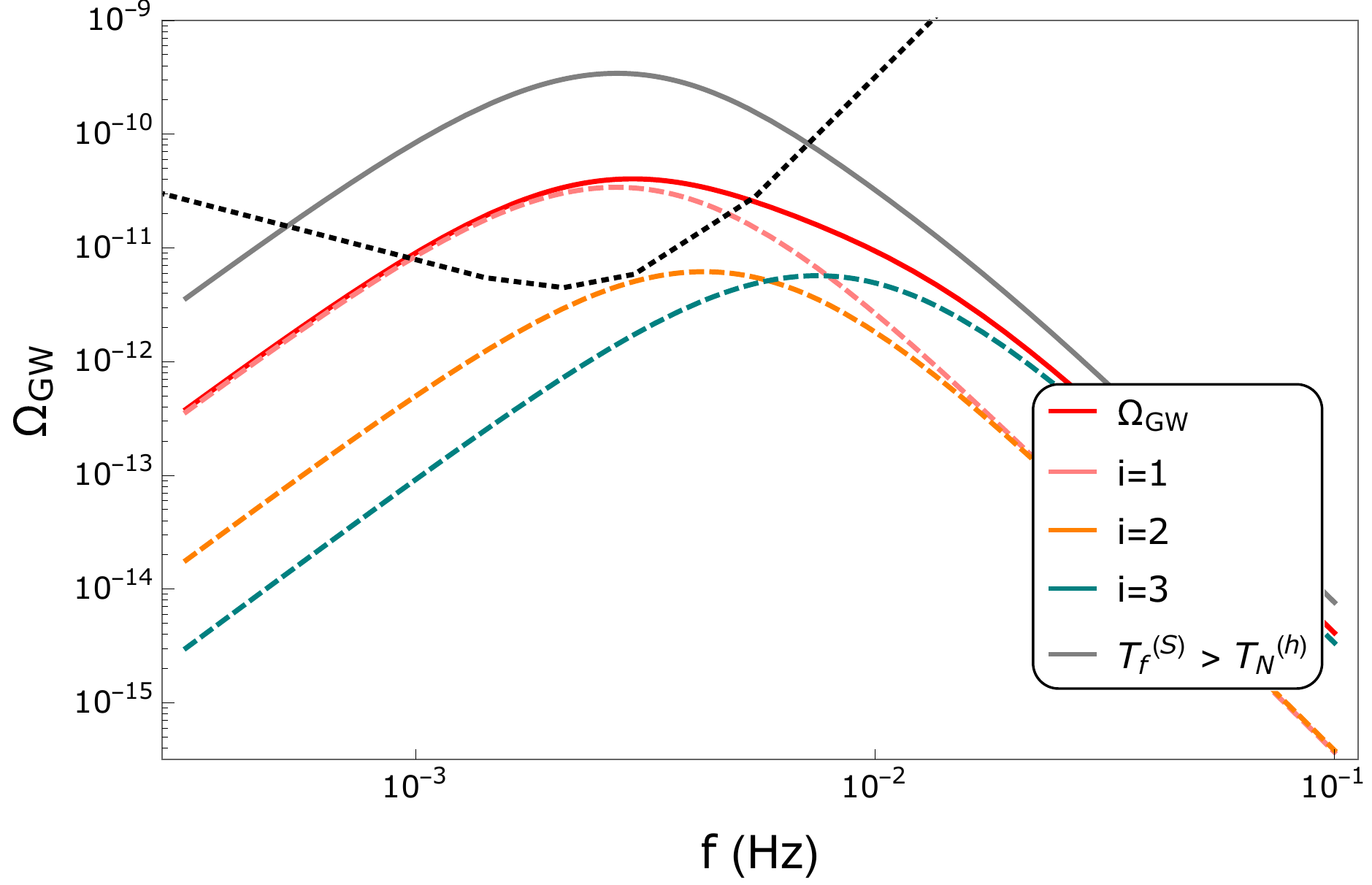}
    \includegraphics[width=.45\textwidth]{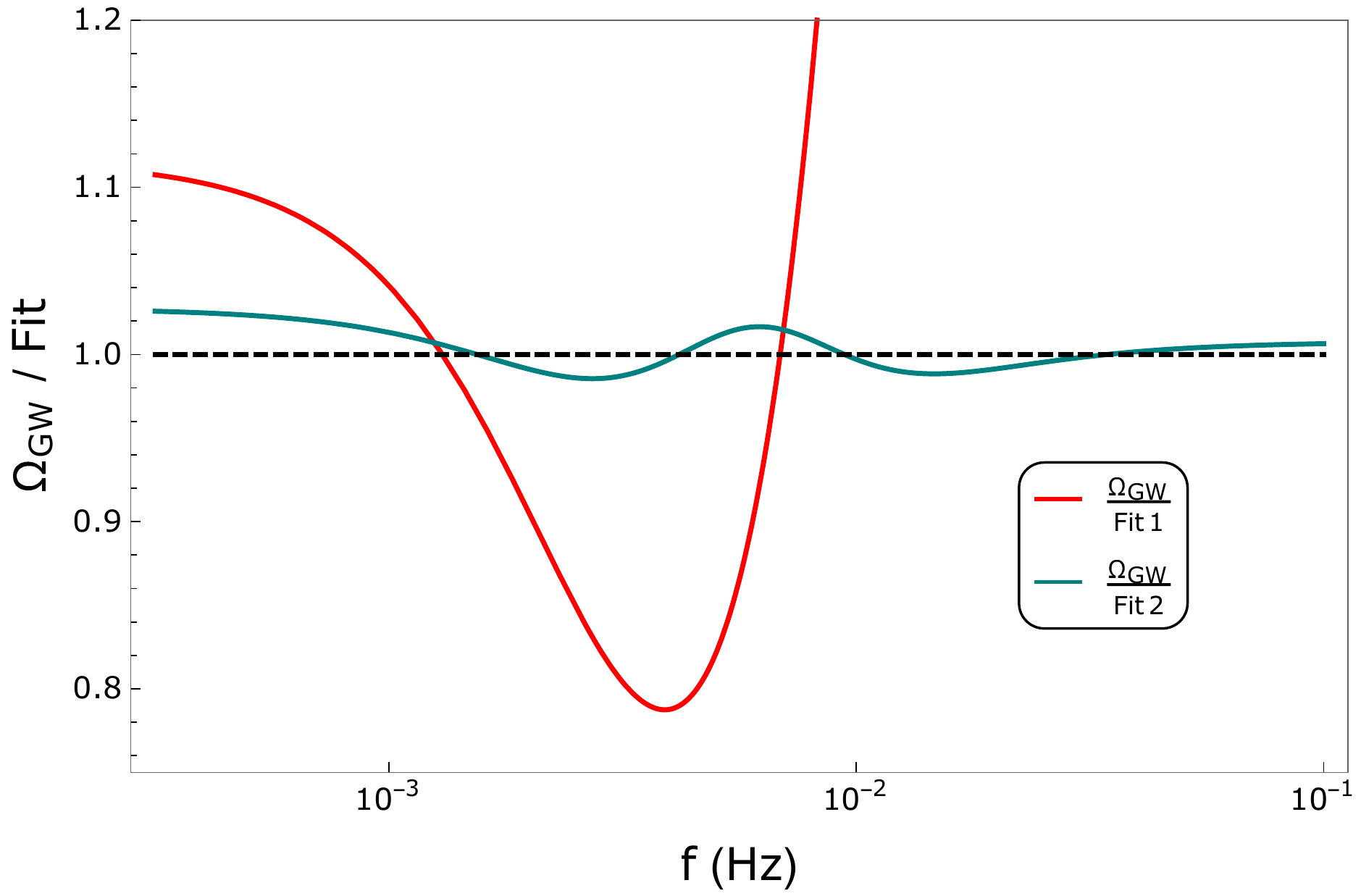}
    \caption{Left panel: the spectrum due to simultaneous singlet and EW PT, given by the red line, leads to a different spectrum than consecutive transitions (gray line). Here we have chosen the second benchmark of Table \ref{tab:benchmwarksGW}, with $w_1=0.1$ and $w_2 = 1- w_3 = 0.15$, and $v^{(2,3)}_w =( 0.6,0.99)$. The dotted line is the projected sensitivity from LISA  \cite{AmaroSeoane:2012km}. Right panel: comparison of a fit to a single and separated phase transition. It is seen that it is not possible to fit to a single peak spectrum. The fit to a doubly peaked spectrum is better, but still cannot mimic the features in the simultaneous bubble case. }
    \label{fig:effectsEW}
\end{figure}

Finally, the gravitational wave spectrum for the third benchmark is given in the left panel of Fig. \ref{fig:effectsEW}. It seen that the overlapping transition differs from the consecutive case, as before. In the right panel of \ref{fig:effectsEW}, we demonstrate once more that the spectrum can not be fitted to that of a single PT. It is seen that in particular, it is not possible to fit both the peak and the tails.

\section{Conclusion and Discussion}
In this work we have considered the gravitational wave signatures from multi-step phase transitions. We have shown that the phase transition can last an extended interval $\Delta T$, and in such cases bubbles of different vacuum phase can coexist. For such scenarios, the resulting gravitational wave signature is expected to be different from multi-step scenarios which are separated in temperature and can be considered consecutively. 

We demonstrated this explicitly by modeling the relative importance of the different nucleation scenarios with weight factors $w_i$. We described a toy model involving two scalar fields (weakly coupled to the Standard Model), and a benchmark scenario of a singlet extension to the Standard Model. While the latter is well motivated by a variety of considerations, especially concerning EW baryogenesis, its detection prospects at future gravitational wave experiments such as LISA are modest. The effects described in this paper may be more relevant for multi-step baryogenesis scenarios with multiple singlets, such as described in \cite{Ramsey-Musolf:2017tgh,Inoue:2015pza}. \par
Throughout this work we have mainly considered the sound wave contribution to the gravitational waves (through efficiency parameters $\kappa_\phi ,\kappa_\text{turb}\ll 1$), as this contribution is expected to dominate \cite{Hindmarsh:2017gnf}.
In this work it was assumed that the curve fitting to numerical simulations holds in our case where the phase transition proceeds very slowly. Furthermore, we have assumed that the thermal parameters are determined by the nucleation scenario; a more sophisticated analysis will allow these to evolve.  
We have also not considered the potentially exotic source of gravitational waves, through the interaction between the plasma shells of bubbles of different species. We leave these effects to future studies.
\begin{acknowledgments}
DC and GW thank Mark Hindmarsh, Daniel Cutting, Chen Sun, Cyril Lagger and David Morrissey for useful discussions.
TRIUMF receives federal funding via a contribution agreement with the National Research Council of Canada and the Natural Science and Engineering Research Council of Canada.
\end{acknowledgments}
\bibliographystyle{JHEP}
\bibliography{references}

\end{document}